\documentclass[journal]{IEEEtran}
\IEEEoverridecommandlockouts

\makeatletter
\def\markboth#1#2{\def\leftmark{\@IEEEcompsoconly{\sffamily}\MakeUppercase{\protect#1}}%
\def\rightmark{\@IEEEcompsoconly{\sffamily}\MakeUppercase{\protect#2}}}
\makeatother

\makeatletter
\newcommand*\titleheader[1]{\gdef\@titleheader{#1}}
\AtBeginDocument{%
  \let\st@red@title\@title
  \def\@title{%
    \bgroup\normalfont\large\centering\@titleheader\par\egroup
    \vskip0.3em\st@red@title}
}
\makeatother

\usepackage{kantlipsum}
\usepackage{setspace}

\usepackage[english]{babel}
\usepackage{xcolor}
\usepackage{lipsum}
\usepackage{caption}
\usepackage{cite}
\usepackage[pdftex]{graphicx}
\usepackage{subcaption}
\usepackage{amsmath}
\usepackage{booktabs}
\usepackage{colortbl}
\usepackage{amsfonts}
\usepackage{amssymb}
\usepackage{amsthm}
\usepackage{array}
\usepackage{verbatim}
\usepackage{listings}
\usepackage{algorithm}
\usepackage{algpseudocode}
\usepackage{url}
\usepackage{enumerate}
\usepackage{multirow}

\definecolor{LightBlue}{rgb}{0.5,0.5,1}
\definecolor{LightRed}{rgb}{1,0.5,0.5}
\definecolor{LightYellow}{rgb}{1,0.85,0}

\usepackage{epsfig}
\usepackage{epstopdf}
\usepackage{multicol}
\usepackage[font=footnotesize]{caption}

\usepackage{algorithm}

\makeatletter
\def\BState{\State\hskip-\ALG@thistlm}
\makeatother

\DeclareMathOperator*{\argmax}{arg\,max}
\renewcommand{\arraystretch}{2}

\newcommand{\bi}{\begin{itemize}}
\newcommand{\ei}{\end{itemize}}
\newcommand{\be}{\begin{equation}}
\newcommand{\ee}{\end{equation}}

\def\beq{\begin{equation}}
\def\eeq{\end{equation}}
\def\beqa{\begin{eqnarray}}
\def\eeqa{\end{eqnarray}}
\def\beqan{\begin{eqnarray*}}
\def\eeqan{\end{eqnarray*}}

\title{Hybrid Spectrum Sharing in mmWave Cellular Networks}
\titleheader{\small M. Rebato, F. Boccardi, M. Mezzavilla, S. Rangan and M. Zorzi, "Hybrid Spectrum Sharing in mmWave Cellular Networks," in IEEE Transactions on Cognitive Communications and Networking, vol. 3, no. 2, pp. 155-168, June 2017.
doi: 10.1109/TCCN.2017.2707551}
\author{{ Mattia Rebato, {Federico Boccardi, \emph{Senior Member, IEEE}\thanks{M. Rebato and M. Zorzi are with the Department of Information Engineering, University of Padova, Italy (e-mail: rebatoma@dei.unipd.it; zorzi@dei.unipd.it).

F. Boccardi is with Ofcom, London SE1 9HA, U.K. (e-mail:
federico.boccardi@ieee.org).

M. Mezzavilla and S. Rangan are with NYU WIRELESS, Tandon School of Engineering, New York University, Brooklyn, NY (e-mail: mezzavilla@nyu.edu; srangan@nyu.edu)}},  Marco Mezzavilla, \emph{Member, IEEE}, Sundeep Rangan, \emph{Fellow, IEEE},  Michele Zorzi, \emph{Fellow, IEEE}\thanks{A preliminary version of this paper was presented at the MED-HOC-NET conference, June 2016~\cite{rebato16hybrid}.}}\vspace{-0.5cm}}

\begin{document}
\maketitle
\begin{abstract}
While spectrum at millimeter wave (mmWave) frequencies is less scarce than at traditional frequencies below 6~GHz, still it is not unlimited, in particular if we consider the requirements from other services using the same band and the need to license mmWave bands to multiple mobile operators. Therefore, an efficient spectrum access scheme is critical to harvest the maximum benefit from emerging mmWave technologies. In this paper, we introduce a new hybrid spectrum access scheme for mmWave networks, where data packets are scheduled through two mmWave carriers with different characteristics. In particular, we consider the case of a hybrid spectrum scheme between a mmWave band with exclusive access and a mmWave band where spectrum is pooled between multiple operators. To the best of our knowledge, this is the first study proposing hybrid spectrum access for mmWave networks and providing a quantitative assessment of its benefits. Our results show that this approach provides advantages with respect to traditional fully licensed or fully pooled spectrum access schemes, though further work is needed to achieve a more complete understanding of both technical and non technical implications.
\vspace{-0.5cm}

\end{abstract}
\bigskip

\section{Introduction}
\label{introduction}

Millimeter wave (mmWave) 
communications are emerging as a key disruptive technology for both cellular networks (5G and beyond)~\cite{pikhan11, ted4, bochea14, andbuz14,sundeep_challenges} and wireless Local Area Networks (802.11ad and beyond)\cite{wigig1,wigig2}.
While spectrum is extremely limited in traditional bands below 6~GHz, mmWave frequencies offer potentially up to two orders of magnitude greater bandwidths.
In addition, thanks to the small wavelength, and due to the need to compensate for the higher path loss at these frequencies, mmWave communications are typically characterized by transmission and reception with very narrow beams, enabling further gains from directional isolation between mobiles~\cite{rohseo14,ted_book}.

This combination of massive bandwidth and spatial degrees of freedom
may make it possible for mmWave to meet some of the boldest 5G requirements, including higher peak per-user data rate, high traffic density, and very low latency~\cite{OssBocc14,alliance20155g}.
However, even the abundant spectrum at mmWave is obviously not unlimited, in particular if we consider the requirements from other services (e.g., satellite and fixed services~\cite{guidolin2015}) and the need to license mmWave bands to multiple mobile operators. Therefore, an efficient spectrum access scheme is critical to harvest the maximum benefit from emerging mmWave technologies~\cite{boccardi16}. 

As  regulatory authorities are considering opening up some mmWave bands for cellular use, the licensing and usage models for these bands require some studies.  
At root, the mmWave bands present three unique features not present at lower frequencies.
First, due to the massive bandwidth and spatial degrees of freedom, the mmWave bands could be highly under-utilized if large bandwidths are allocated exclusively to a single operator.
For example, a scaling law analysis in \cite{felipe} as well as simulations in \cite{mio3} demonstrate that links may become power-limited in wide bandwidth regimes, thereby forgoing the benefits of the large numbers of degrees of freedom.
Second, mmWave communications are typically characterized by transmissions with very narrow beams.
Third, mmWave signals suffer from major propagation-related shortcomings, such as a relatively short range and the difficulty of providing a robust connection, which makes it challenging to provide a consistent user experience. 
To overcome these shortcomings, mmWave networks have been usually envisioned
in the context of heterogeneous deployments \cite{ghotho14, and13, IshKis12, ElsKul16}, where part of the connection is carried out with an anchor over a traditional sub-6 GHz carrier and part via a mmWave carrier.  

More recent results have shown that even stand-alone mmWave systems can be deployed, and in this case it becomes of interest to study systems where the 
use of different bands in the very wide mmWave spectrum (e.g., at 
lower frequencies -- around 30 GHz -- and at higher frequencies -- around 70 GHz) may provide complementary features, thereby enabling a more efficient use of the spectrum resources, especially in the context of a spectrum sharing paradigm~\cite{boccardi16}. Such a 
\emph{heterogeneous mmWave deployment} paradigm is consistent with the choice made by the 2015 World Radio Conference, where different bands, ranging from about 24~GHz to 86~GHz, were selected for further studies on their use in future 5G systems~\cite{final_acts_wrc_15}.

These features raise some broad questions that are the main motivation for this paper, e.g., how the mmWave spectrum should be utilized amongst multiple operators and, 
specifically, to what extent spectrum should be shared and how 
the optimal spectrum sharing arrangement varies with the different frequency
bands. The  main goal of this paper is to provide some initial answers to
these important questions, with focus on technical and network
performance issues.\footnote{We recognize that economic factors will also require study, and 
that a more complete picture will eventually need to include both.}

\subsection{Traditional spectrum access models for mobile communications}
Traditionally, wireless data services have been delivered mainly by using two different spectrum access models.
Under the \textit{exclusive} model, each mobile operator is granted the exclusive right of use of a spectrum band to provide mobile services.
Exclusive spectrum access has been one of the key factors for the successful deployment of cellular systems since their inception, and it is by far the default model to provide mobile services.
Under the \textit{license-exempt} (also referred to as  \textit{unlicensed}) model, spectrum is allowed to be used by several users/mobile operators.
While there is no guaranteed access to an instantaneously fixed amount of spectrum, politeness rules (e.g., based on a listen-before-talk principle) are in place to promote a fair use of the spectrum.
The license-exempt spectrum model has been one of the key factors for the successful deployment of WiFi as a ubiquitous way of connecting devices to the Internet.
The \textit{spectrum pooling} model has also been considered as an intermediate paradigm, where a small number of operators are granted access to the same spectrum resources, with rules that are known a priori~\cite{Jondral04}.
Although spectrum pooling still does not provide guarantees for the access to an instantaneously fixed amount of spectrum, it does ensure some level of predictability and of short-term and long-term fairness~\cite{METIS51}.
We note that spectrum pooling is a subcase of co-primary spectrum sharing, where an operator is authorized to share a band with a limited number of other spectrum users (for example sharing between fixed links and satellite services licensed on the same band)~\cite{METIS51,Irnich13}.
Many other sharing paradigms have been considered in addition to co-primary sharing, like sharing between a primary and a secondary user (vertical sharing)~\cite{gupta_jsac_2016,safavi15}, sharing on a geographic basis~\cite{Zheng14}, licensed shared access (e.g., via databases)~\cite{matinmikko14}, and sharing via license-exemption~\cite{Bhattarai}. However, herein we mainly consider spectrum pooling, although our results can be extended also to the case of license-exemption.  

\subsection{The emergence of hybrid spectrum access for sub-6~GHz wireless communications}
Recently, new technologies have emerged that aggregate spectrum in both exclusive and license-exempt bands, routing packets to the carrier frequency that best matches their requirements.
Aggregation is implemented in a way to permit a very rapid switch between exclusive and license-exempt carriers, effectively realizing a \textit{hybrid spectrum access regime}.
Examples of these technologies are Long Term Evolution-Licensed Assisted Access (LTE-LAA), LTE-WiFi Link Aggregation (LWA) and LTE-WiFi integration at the IP layer (LWIP)~\cite{laa1,laa2,lte_wifi,liason_statement,Galanopoulos16}.
LAA is an extension of carrier aggregation that allows aggregating licensed carriers with license-exempt spectrum at 5~GHz, in the same bands used for WiFi.
In particular, it uses licensed spectrum for control-related transmissions while sending data over either licensed or licence-exempt spectrum via MAC-layer switching\footnote{An equitable coexistence between LAA and WiFi is guaranteed by mandating that both LAA and WiFi implement a set of politeness protocols, whose details have been recently defined by the European Telecommunications Standards Institute (ETSI) Broadband Radio Access Networks (BRAN) committee.}.
LWA is a framework standardized by 3GPP aiming at providing a tight radio-level interaction between LTE and WiFi.
Using LWA, aggregation between LTE and WiFi is implemented at the base station at the PDCP layer, where scheduling decisions can be made based on real-time channel conditions.
LWIP is similar to LWA, but aggregates traffic at the IP layer in a way to route IP packets to either an LTE base station or a WiFi access point via an IPSec tunnel.

\subsection{Which spectrum access for mmWave networks?}
As discussed above, an efficient spectrum access scheme is a key requirement to maximally benefit from emerging mmWave technologies~\cite{boccardi16}. Recent works compared exclusive spectrum allocation with different types of spectrum pooling or unlicensed models, showing different results as a function of the assumptions used. Reference~\cite{li2014} introduced a new signaling report among mobile operators, to establish an interference database to support scheduling decisions, with both a centralized and a distributed supporting architecture. In the centralized case, a new architectural entity receives information about the interference measured by each network and determines which links cannot be scheduled simultaneously. In the decentralized case, the victim network sends a message to the interfering network with a proposed coordination pattern. The two networks can further refine the coordination pattern via multiple stages. Reference~\cite{gupta15} studied the feasibility of spectrum pooling in mmWave networks under the assumption of ideal antenna patterns and showed that spectrum pooling might be beneficial even without any coordination between the different operators. In particular,~\cite{gupta15} showed that uncoordinated pooling provides gains at both 28~GHz and 73~GHz.  Reference~\cite{boccardi16} further developed the results in~\cite{li2014} and~\cite{gupta15},  focusing on the effect of coordination and of inaccurate beamforming, and showed that, while coordination may not be needed under ideal assumptions, it does provide substantial gains when considering more realistic channel and interference models and antenna patterns. Moreover,  it showed that, under realistic assumptions, spectrum pooling without coordination might be more feasible at high mmWave frequencies (e.g., 70~GHz) than at low mmWave frequencies (e.g., 28 or 32~GHz), due to the higher directionality of the beams.
Reference~\cite{rebato16} compares different resource sharing paradigms and shows that a full spectrum and infrastructure sharing configuration provides significant advantages, even without resorting to complex signaling protocols for the exchange of information between multiple operators' networks.
Reference~\cite{hossein16} investigates the use of spectrum sharing as a function of cell association and beamforming, through the formulation of various optimization problems for different levels of inter-operator coordination, and characterizes the performance gains achievable in different scenarios. 
Finally, reference~\cite{fund16} studies both technical and economic implications of resource sharing in mmWave networks.
The work shows that open deployments of neutral small cells that serve subscribers of any service provider encourage market entry.
In fact, neutral small cells make it easier for networks to reach a sufficient number of subscribers than unlicensed spectrum would. 

\subsection{The contribution of this paper: hybrid spectrum access for mmWave networks}

\begin{figure}[t!]
\centering
\includegraphics[width=0.8\columnwidth]{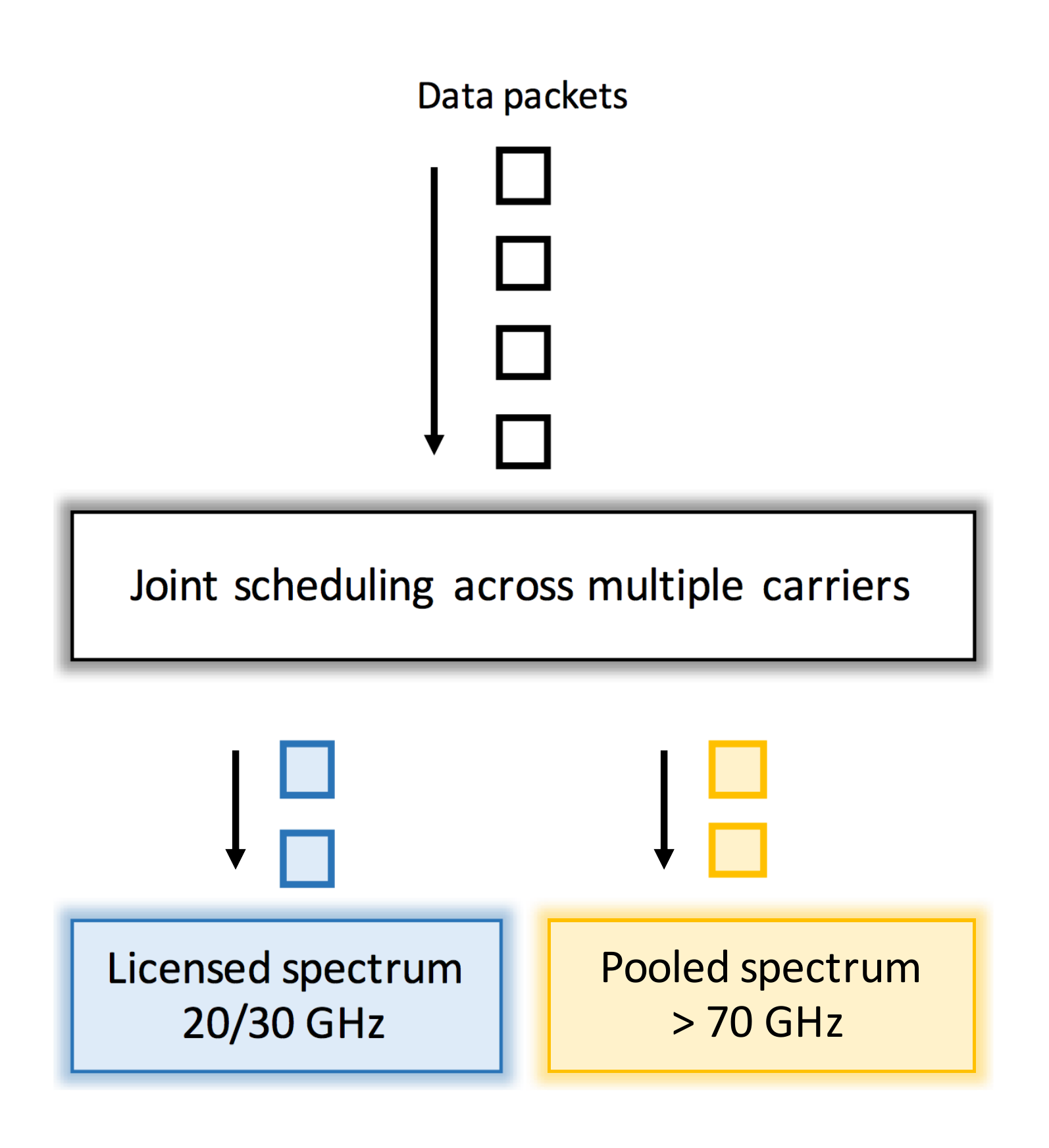}
\caption{Block diagram of the joint scheduling that allocates user packets in the different bands.}
\label{block_scheme}
\end{figure}

This paper extends the previous results in~\cite{rebato16hybrid,boccardi16,rebato16} and~\cite{hossein16} to the case of hybrid spectrum allocation. In other words, differently from the previous works, where exclusive access and spectrum pooling were compared, in this work we propose a spectrum access paradigm that builds on both exclusive access and spectrum pooling.
We introduce the use of an iterative algorithm to evaluate the equilibrium point of the system, and therefore to precisely appraise our hybrid spectrum sharing procedure.
Note that this algorithm is not meant to represent how a real system would work but is just one possible tool to evaluate the system.
Moreover, with the use of this algorithm we can easily test different allocation procedures (such as joint carrier and cell or carrier-only), thus we can further show the benefit of the hybrid procedure suggested under different parameters and different power constraints.
 In particular, motivated by the results in~\cite{boccardi16} where pooling was proved to be more feasible at high mmWave frequencies, we  study the performance of a hybrid spectrum scheme where exclusive access is used at frequencies in the 20/30~GHz range while pooling is used at frequencies around 70~GHz\footnote{In the following we will refer to the 28~GHz and 73~GHz bands, for which many measurements are available in the literature (e.g., see~\cite{mustafa,ted1,ted2,ted3,ted4}).
However, we note that the results herein, possibly with some minor modifications, would apply to adjacent bands as well.
In particular, the results obtained for the 28~GHz band apply also to the two bands selected by the 2015 World Radio Conference (WRC-15)~\cite{final_acts_wrc_15} for sharing and compatibility studies for 5G, i.e., 24.25~--~27~GHz and 31.8~--~33.5~GHz.
The results obtained for 73~GHz apply to the 66~--~76~GHz band, again selected by WRC-15 for sharing and compatibility studies for 5G.}.
The two bands are aggregated at the MAC layer as illustrated in Figure~\ref{block_scheme}, and users are allocated to one or the other band to maximize the rate, based on cell load and interference.

Differently from the LAA case, in this study we assume that all the operators sharing the pooled band have access to an anchor in the licensed spectrum. Moreover, differently from LAA, that includes politeness techniques based on a listen-before-talk protocol permitting coexistence with WiFi within the same 5~GHz bands, here we investigate the possibility of providing politeness between different operators sharing the pooled mmWave band by exploiting mmWave directional characteristics (narrower beams and shorter range) through load information and an interference-aware scheduler\footnote{Note that with this term we generically indicate a scheduler that is able to choose the less interfered carrier. However, this work does not aim to analyze in depth any particular kind of scheduler.}.

Finally, we note that we consider aggregation between a licensed and a pooled carrier rather that between a licensed and a license-exempt carrier.
Overall, the different spectrum sharing assumptions (pooled and unlicensed, with and without an anchor in licensed spectrum) and the very different directional characteristics lead us to designing a different solution for hybrid spectrum access at mmWave, compared to the solutions already available for sub-6 GHz spectrum.

We compare our proposal with two baselines, one relying on exclusive spectrum access at both 28~GHz and 73~GHz and the other relying on pooling at both 28~GHz and 73~GHz.
To the best of our knowledge, this is the first study providing evidence on the benefits of hybrid spectrum access for mmWave networks.
Our initial assessment shows that this approach provides advantages for the average user with respect to traditional fully licensed or fully pooled spectrum access schemes, in terms of increased throughput, spectral efficiency, and better balancing of the available resources, which results in higher fairness.
These results motivate further work towards achieving a more complete understanding of both technical and non-technical implications of different sharing paradigms.

The rest of the paper is organized as follows.
In Section \ref{hybrid_spectrum_access}, we describe the proposed hybrid spectrum allocation scheme and the two baselines we will use for comparison, and
in Section \ref{simulation_methodology} we provide the simulation methodology used for this study.
In Section \ref{association}, we introduce our base station and carrier association algorithms, and in Section \ref{simulation_results} we present a numerical evaluation and discuss the results.
Finally,  we describe some future research steps and conclude the paper in Section~\ref{futureworks}.

\section{Spectrum Access Modes:  Exclusive, Pooled and Hybrid}
\label{hybrid_spectrum_access}
\begin{figure*}[t!]
\centering
\includegraphics[width=0.67\textwidth]{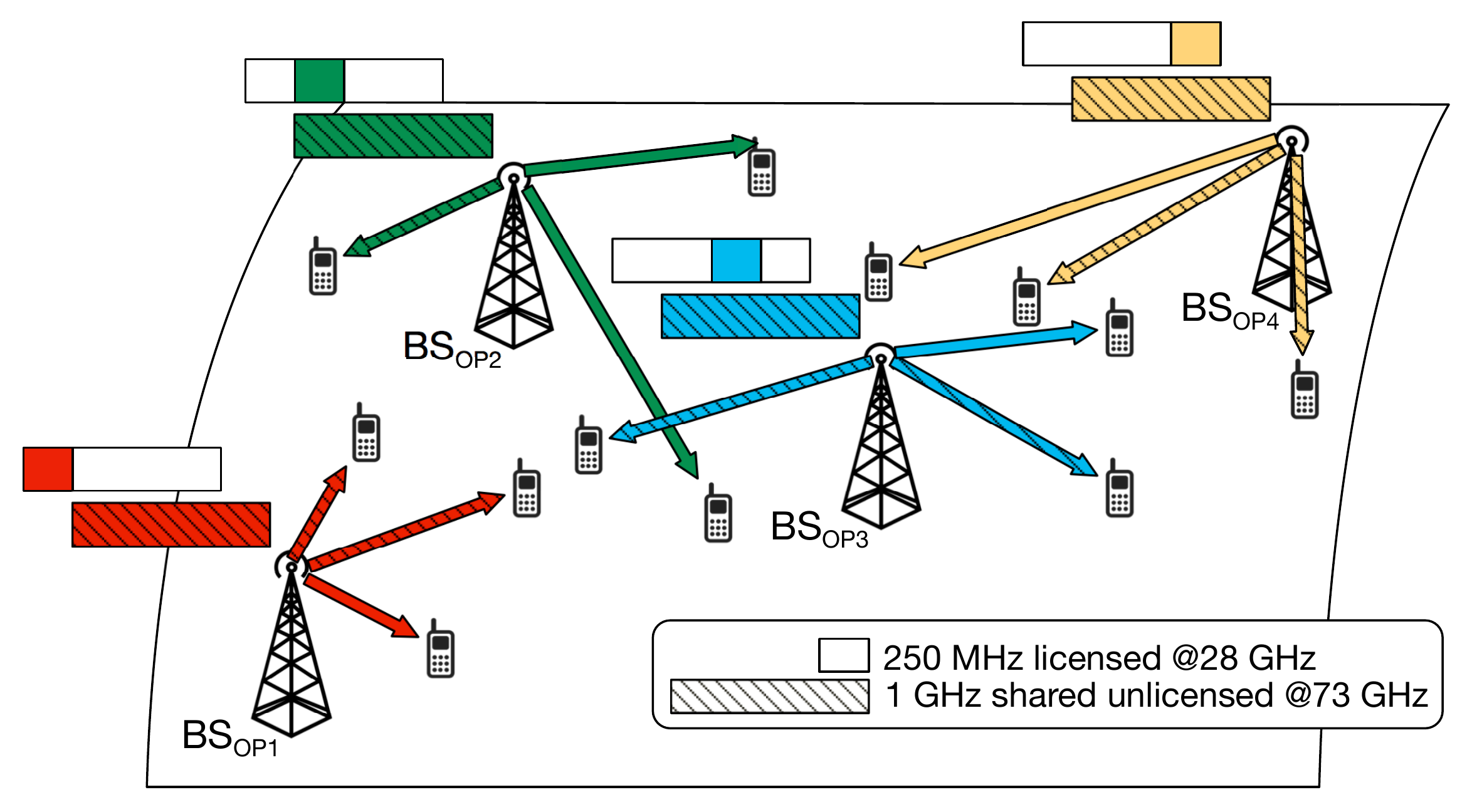}
\caption{Example of the hybrid spectrum paradigm: four operators share 1~GHz in the 70~GHz range, while having each exclusive access to 250~MHz in the 20/30~GHz range.}
\label{scenario_draw}
\end{figure*}

We begin by more precisely defining the various modes for spectrum access.
We consider a scenario with $M$ operators indexed by $m \in \{1,\ldots,M\}$.  
Each operator owns distinct base stations (BSs) with no infrastructure sharing between 
operators.
Each BS supports two mmWave bands: one at a low carrier $c_{\ell}$ and one at a high carrier $c_h$.
In each carrier $c$, a BS can radiate a maximum total power $P_{\text{TX}}^{(c)}$ over the available 
bandwidth $W^{(c)}$.

Each carrier can be pooled or exclusive. Let $W_{tot}^{(c)}$ be the total system bandwidth 
available at carrier $c$.
Exclusive access means that each operator is assigned a bandwidth $W^{(c)}=W_{tot}^{(c)}/M$, such that bands assigned to different operators are disjoint.  As a result, in the exclusive case, there is no co-channel interference between different operators.
A carrier being pooled instead means that it is shared by all $M$ operators, so that in this case all operators use the same bandwidth and
$W^{(c)}=W_{tot}^{(c)}$. Hence, in this case there is co-channel interference between operators.  

In this paper, we propose and evaluate the following novel hybrid spectrum access model for systems with two carriers:
\begin{itemize}
\item \emph{Hybrid:}  The low frequency carrier $c_\ell$ is exclusive while the high frequency carrier $c_h$ is pooled (see Fig.~\ref{scenario_draw} for an example),
\end{itemize}
which will be compared with the following two baseline approaches:
\begin{itemize}
\item \emph{Exclusive only:}  Both carriers are licensed for exclusive use for all operators;
\item \emph{Pooled only:}  Both carriers are pooled for all operators.
\end{itemize}

The use of the hybrid access scheme is motivated by the results in~\cite{boccardi16} that showed that while at  ``low'' $c_{\ell}$ mmWave frequencies (e.g., 20/30~GHz) spectrum pooling requires coordination between different operators, at higher $c_h$ mmWave frequencies (e.g., 70~GHz) pooling works well even in an uncoordinated setup.
As an example of a possible use of this paradigm, data requiring higher reliability (e.g., control signaling) is routed through exclusive mmWave spectrum while best-effort data uses pooled mmWave spectrum.

We note that this approach is reminiscent of the LAA framework, recently standardized by 3GPP~\cite{laa1}.
LAA is an extension of carrier aggregation in which licensed carriers are aggregated with license-exempt spectrum at 5~GHz.
The main differences between LAA and the approach we propose is that here we consider aggregation between a licensed and a pooled carrier rather than between a licensed and a license-exempt carrier\footnote{We highlight that the main difference between pooled and license-exempt frequencies is that if we enable a licence-exempt use of the band, we ought to consider mechanisms to ensure an equitable use of the spectrum.
From a technical perspective, this would require further steps compared to what we have proposed up to now. At lower frequencies (5~GHz) this is already done, but mechanisms used there (e.g., listen-before-talk) might not apply or might not be optimized for mmWave frequencies.}.
Moreover, differently from LAA, that includes politeness techniques to allow coexistence with WiFi within the same 5~GHz bands, here we investigate the possibility of aggregating licensed and pooled carriers, by exploiting the directional characteristics (narrower beams and shorter range) of mmWave bands via an interference-aware scheduler.

From a spectrum authorization perspective, the difference between LAA and the scheme we propose here is that LAA is built to work on a shared band that is license-exempt, i.e., where everyone can have an access point compliant with the RLAN standard and deploy it.
Some of these access points might be LAA access points (and therefore exploit an exclusively licensed carrier in a different frequency band), while others might be WiFi access points (and therefore only use license-exempt spectrum).
The way regulators ensure that LAA and WiFi access points equitably share the spectrum is by mandating the use of a set of politeness protocols.
In this paper we consider the case where the spectrum is shared (pooled) by a limited number of users, all of which have access to an exclusively-licensed band at a different frequency.
From a technical perspective, there is a significant difference related to the use of politeness protocols.
Due to the specific authorization assumption we make, we do not consider politeness protocols, and design our proposed user/carrier allocation technique accordingly.
In our future works we plan to extend this to the case of license-exempt spectrum, and we will investigate the required politeness protocols.
We note that the politeness protocols used for LAA/WiFi at 5~GHz have been specifically designed to account for the characteristics of the interference at traditional sub-6~GHz band.
New studies must be done to understand how politeness protocols should be designed for cellular communications at mmWave.
We will further discuss this topic in Section \ref{futureworks}.

\section{Evaluation Methodology}
\label{simulation_methodology}

A precise mathematical analysis of the capacity under different spectrum access models is 
difficult due to the inter-relations among interfering operators, the coupling introduced by load-aware association policies, and the complex characterization of the multiple-input multiple-output (MIMO) channel model.
Such an analysis requires careful modeling and the use of approximations, and will be part of our future work.
Here, in order to provide a proof-of-concept evaluation for the proposed hybrid spectrum access approach under realistic scenarios for mmWave cellular systems, we study the different spectrum access schemes through a careful simulation methodology, where detailed models are used for all important effects and variables (including in particular channel characteristics and association policies), as described below.

\paragraph*{Deployment model}
For each operator $m \in \{1,\ldots,M\}$, 
the positions of the user equipments (UEs) and of the BSs are modeled according to two Poisson Point Processes (PPPs), with densities $\lambda_{\text{UE}}$ and $\lambda_{\text{BS}}$ in some area $A$.
This corresponds to considering an unplanned deployment, where base stations are not optimally located.


\paragraph*{Rate and scheduling model}
We let $\mathbf{H}^{(c)}_{ij}$ denote the MIMO channel matrix 
from BS $i$ to UE $j$ at carrier $c$.  For simplicity, in this initial study we assume that the channel gain is flat
across time and frequency.  
We assume beamforming with single-stream transmissions (i.e., no spatial multiplexing) to any one UE.
We let $\mathbf{w}_{\text{RX}_{ij}}^{(c)}$ and $\mathbf{w}_{\text{TX}_{ij}}^{(c)}$ denote the RX and TX beamforming vectors that would be used if BS $i$ were serving UE $j$.
The generation of the channel matrices and selection of the beamforming vectors is discussed below.

With single-stream 
beamforming, the effective single-input single-output (SISO) channel gain along the serving link is given by:
\begin{equation} \label{Gmij_eq}
    G_{ij}^{(c)} = \left| \mathbf{w}_{\text{RX}_{ij}}^{(c)H}\mathbf{H}^{(c)}_{ij}\mathbf{w}_{\text{TX}_{ij}}^{(c)}\right|^2.
\end{equation}
Now, consider the gain from an interfering BS $k$.  An interfering BS may be from 
the same operator as that of the UE or a different operator within a common pooled band.
In either case, the UE will experience a time-varying interference as the interfering BS
directs its transmissions to the different UEs it is serving.  We let $\bar{G}^{(c)}_{ijk}$ be the average
channel gain from the interfering BS $k$ to user $j$ of BS $i$, defined as:
\begin{equation} \label{Gbar_eq}
    \bar{G}^{(c)}_{ijk} = \frac{1}{N^{(c)}_{k}} \sum_{j'} 
        \left| \mathbf{w}_{\text{RX}_{ij}}^{(c)H}\mathbf{H}^{(c)}_{kj}\mathbf{w}_{\text{TX}_{kj'}}^{(c)}\right|^2,
\end{equation}
where the averaging is over all UEs $j'$ served by BS $k$.
The Signal-to-Interference-plus-Noise Ratio (SINR) is then given by:
\begin{equation}
\gamma_{ij}^{(c)} = \frac{\frac{P^{(c)}_{\text{TX}_{i}}}{PL_{ij}^{(c)}}G^{(c)}_{ij}}
{\sum_{k \neq i}  \frac{P^{(c)}_{\text{TX}_{k}}}{PL^{(c)}_{kj}}\bar{G}^{(c)}_{ijk} + W^{(c)}  N_0},
\label{equation_sinr}
\end{equation}
where $P^{(c)}_{\text{TX}_i}$ is the total transmit power from the BS in the available bandwidth $W^{(c)}$ at carrier $c$, $N_0$ is the thermal noise power spectral density and $PL_{ij}^{(c)}$ is the path loss between BS $i$ and UE $j$ and is computed as described in the following paragraphs.
The summation in the denominator of~\eqref{equation_sinr} is over all BSs $k$ in the band, including BSs of both the same operator and other operators.
Note that, within the cell, we assume that UEs are scheduled on orthogonal resources (e.g., in time or frequency) and hence there is no intra-cell interference.

\paragraph*{MIMO Channel Model}
The MIMO channel matrices are generated according to a statistical channel model
derived from a set of extensive measurement campaigns in New York City
\cite{ted1,ted2,ted3,mathew}.  Details of this model are given in  \cite{mustafa}.  Briefly, 
each link is independently determined to be in one of three states: line-of-sight (LOS),
non-line-of-sight (NLOS) or outage (out) with a probability that depends on the link distance $d$ (see Table~\ref{table_pathloss_probabilities}).
\begin{table}[h!]
\normalsize
\renewcommand{\arraystretch}{1.15}
\centering
\begin{tabular}{c|c}
\toprule
\emph{out} & $p_{\mathrm{out}}(d)=\mathrm{max}(0,1-e^{-a_{\mathrm{out}}d+b_{\mathrm{out}}})$\\ \hline
LOS & $p_{\mathrm{LOS}}(d)= (1-p_{\mathrm{out}}(d))e^{-a_{\mathrm{LOS}}d}$ \\ \hline
NLOS & $p_{\mathrm{NLOS}}(d)= 1-p_{\mathrm{out}}(d)-p_{\mathrm{LOS}}(d)$ \\ 
\bottomrule
\end{tabular}
\caption{Path loss model functions to compute the probability to be in one of the three states, where $a_{\mathrm{out}}=0.0334$ m$^{-1}$, $b_{\mathrm{out}}=5.2$ and $a_{\mathrm{LOS}}=0.0149$ m$^{-1}$ (all these values are taken from~\cite{mustafa}).}
\label{table_pathloss_probabilities}
\end{table}
For links in outage, there is no connection between the BS and the UE.  For LOS and NLOS links, 
the omni-directional path loss follows a distance-dependent attenuation given by:
\be
PL(d)[dB] = \alpha + \beta 10 \log_{10}(d) + \xi,
\label{pathloss_equation}
\ee
where $\xi\sim N(0,\sigma^2)$ is the log-normal shadowing, and the parameters $\alpha,\beta,\sigma$,
derived in~\cite{mustafa}, are reported in Table~\ref{table_pathloss_parameters} and depend on the carrier and on the LOS or NLOS link state. 

\begin{table}[h!]
\normalsize
\renewcommand{\arraystretch}{1.15}
\centering
\begin{tabular}{c|c|c|c|c}
\bottomrule
Frequency & State & $\alpha$ & $\beta$ & $\sigma$ \\ \hline \hline
\multicolumn{1}{c|}{\multirow{2}{*}{28~GHz}} & NLOS & 72 & 2.9 & 8.7~dB   \\ \cline{2-5}  
\multicolumn{1}{c|}{}                        & LOS  &  61.4  &  2   &   5.8~dB       \\ \hline
\multirow{2}{*}{73~GHz}                     &  NLOS    & 86.6   &  2.45   &   8.0~dB        \\ \cline{2-5}  
                                            &   LOS   & 69.8  &   2  &     5.8~dB                            \\ [-2pt]\bottomrule
\end{tabular}
\caption{Path loss model parameters derived from real measurements made in NYC~\cite{mustafa}. Values for both the 28 and 73~GHz bands, and for both LOS and NLOS conditions.}
\label{table_pathloss_parameters}
\end{table}

We consider a wrap-around procedure that replicates each transmitting BS in eight different additional areas around the main area. With this method, we remove the cell-edge effects by considering all the interfering terms, thereby correctly evaluating the statistics of interest for all the users.

The channel is also composed of a random number $K$ of clusters with random angles, angular spread
and power.  The models for the large-scale parameters are also carrier and link-state dependent.

We model the antennas as a uniform planar array (UPA) with $\lambda/2$ spacing at both the BS and the UE.
Once the large-scale parameters are randomly generated, a random matrix $\mathbf{H}^{(c)}_{ij}$ can be 
generated from the UPA array and random small-scale complex fading applied to each sub-path in the path cluster.
Further details are in \cite{mustafa}.
\begin{figure*}[t!]
\begin{equation}
\footnotesize
\textbf{w}^{(c)}_{\text{TX}_{mij}}(\theta,\phi) = \frac{1}{\sqrt {n_{\text{TX}}}} \begin{bmatrix} 1\\ \exp(0) \exp(-j2\pi \Delta \phi)\\ \exp(0) \exp(-j2\pi \Delta 2 \phi) \\ \vdots \\ \exp \left(-j2\pi (\sqrt{n_{\text{TX}}}-1) \Delta \theta \right) \exp(-j2\pi (\sqrt{n_{\text{TX}}}-2) \Delta \phi) \\ \exp \left(-j2\pi (\sqrt{n_{\text{TX}}}-1) \Delta \theta \right) \exp(-j2\pi (\sqrt{n_{\text{TX}}}-1) \Delta \phi)\end{bmatrix}, 
\normalsize
\end{equation}
\rule[0.5ex]{\linewidth}{1pt}
\end{figure*}

\paragraph*{Beamforming}  
For each channel matrix $\mathbf{H}^{(c)}_{ij}$ we compute the BF vectors at the transmitter (or receiver) as reported at the top of this page, where $\Delta$ is the spacing between the elements of the array, $(\theta,\phi)$ are the horizontal and vertical angles of the direction of transmission (or reception in the RX case), and $n_{\text{TX}}$ is the normalization factor, which corresponds to the total number of elements in the antenna array~\cite{antenna_book}.
Note that the only difference between RX and TX is the number of antenna elements.
Among other simplifications, this model
assumes perfect beam tracking and the ability to form an arbitrary BF vector.
Therefore, we can generate a beamforming vector for any possible angle between 0 and 360 degrees. We also assume perfect alignment between the beams of each UE and its serving BS.

\paragraph*{Antenna configuration and Power Limits}
We consider three different transmitter and receiver configurations:
\begin{description}
\item [\emph{i)}] Both bands use the same number of antenna elements: $n_{\text{TX}} = 64$ and  $n_{\text{RX}} = 16$, and are subject to the same constraint on the Equivalent Isotropically Radiated Power (EIRP), i.e.:
\begin{equation}
E\left[ \vert {\textbf{x}^{(c)}}^H_{mij}{\textbf{x}^{(c)}}_{mij} \vert ^2 \right] \leq P_{\text{TX}},\quad c \in \lbrace c_{\ell},c_h \rbrace,
\end{equation}
where $\textbf{x}^{(c)}_{mij}$ is the symbol exchanged between BS $i$ and UE $j$ of operator $m$ using carrier $c$.

\item [\emph{ii)}] We double the number of antenna elements per dimension for the higher band\footnote{For example, due to the reduced wavelength, with the same antenna array dimension at 73~GHz we can fit about 2.6 times more elements per dimension than at 28~GHz.}. Moreover,  we normalize the beamforming coefficients in a way to satisfy the same EIRP constraint for both bands as in \emph{i)}.

\item [\emph{iii)}] As in \emph{ii)}, we double the number of antenna elements per dimension for the higher band. However, we consider different EIRP constraints for the different bands:
\begin{equation}
E\left[ \vert {\textbf{x}^{(c)}}^H_{mij}{\textbf{x}^{(c)}}_{mij} \vert ^2 \right] \leq  P^{(c)}_{\text{TX}},\quad c \in \lbrace c_{\ell},c_h \rbrace.
\end{equation}

Under the assumption of $n_{\text{TX}} = 64$ and  $n_{\text{RX}} = 16$ for $c_{\ell}$ and $n_{\text{TX}} = 256$ and  $n_{\text{RX}} = 64$ for $c_h$, we can assume the following values for the antenna array gains:

\begin{equation}
\centering
\begin{split}
G^{(c_{\ell})} &= 10 \log_{10} (64) \simeq  18\, \text{dB},\\
G^{(c_h)} &= 10 \log_{10} (265) \simeq  24\, \text{dB}.
\label{max_gains}
\end{split}
\end{equation}

Then:

\begin{equation}
\begin{aligned}
P^{(c_h)}_{\text{TX}} [dB] &=   P^{(c_{\ell})}_{\text{TX}} [dB] + G^{(c_h)}-G^{(c_{\ell})}\\
&=P^{(c_{\ell})}_{\text{TX}} [dB] + 6.
\end{aligned}
\end{equation}

\end{description}

\begin{table*}[h!]
\normalsize
\renewcommand{\arraystretch}{1.3}
\centering
\begin{tabular}{r|c|c|c|c|cc}
\toprule
\multicolumn{1}{l|}{} & \multicolumn{2}{c|}{\begin{tabular}[c]{@{}c@{}}\# UE antenna\\ elements\end{tabular}} & \multicolumn{2}{c|}{\begin{tabular}[c]{@{}c@{}}\# BS antenna\\ elements\end{tabular}} & \multicolumn{2}{c}{\begin{tabular}[c]{@{}c@{}}BS Power limit\\ $P_{\text{TX}}$ {[}dBm{]}\end{tabular}} \\ \cline{2-7} 
\multicolumn{1}{l|}{} & 28 GHz               & 73 GHz               & 28 GHz               & 73 GHz               & \multicolumn{1}{c|}{28 GHz}       & 73 GHz       \\ \toprule \bottomrule
\textit{i)}           & 16                   & 16                   & 64                   & 64                   & \multicolumn{1}{c|}{30}           & 30           \\ \hline
\textit{ii)}          & 16                   & 64                   & 64                   & 256                  & \multicolumn{1}{c|}{30}           & 24           \\ \hline
\textit{iii)}         & 16                   & 64                   & 64                   & 256                  & \multicolumn{1}{c|}{30}           & 30    \\
\bottomrule      
\end{tabular}
\caption{Antenna element sizes and transmit power limits for the various power constraints.}
\label{table_power_contraint_configuration}
\end{table*}

\begin{figure*}[t!]
\centering
\includegraphics[width=\textwidth]{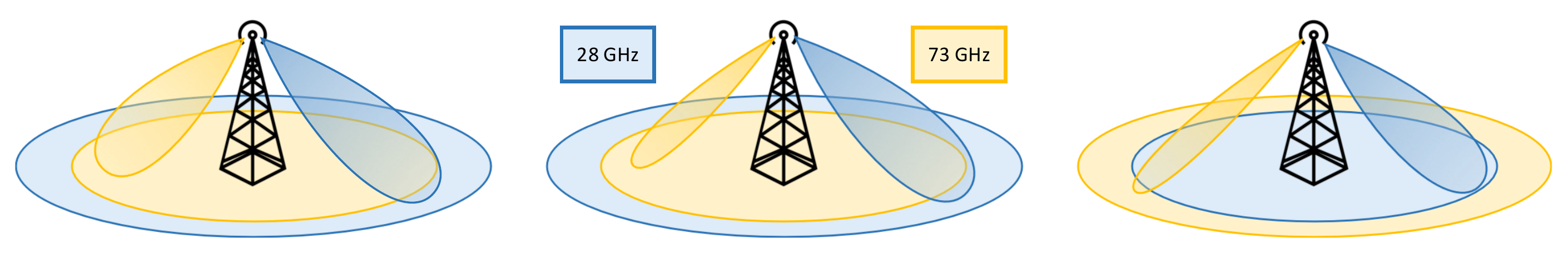}
\caption{Examples of beamforming and coverage in the two bands for the three power constraint scenarios. The left drawing is case \emph{i)}, the one in the middle is constraint \emph{ii)} and finally, the right drawing is case \emph{iii)}.}
\label{pow_budgets}
\end{figure*}

We provide a graphical representation of the implications of the different configurations on the transmit beams in Figure~\ref{pow_budgets}, and summarize the parameters for the different configurations in Table~\ref{table_power_contraint_configuration}.
Configuration \emph{i)} implies that the same beamwidth is used for the two bands.
Moreover, due to the increased path loss at higher frequencies, the higher band provides a reduced coverage area.
Configuration \emph{ii)} implies that a narrower beam is used for the higher frequency.
However, EIRP normalization at the higher frequency implies that also in this case the higher frequency might provide a reduced coverage area.
Configuration \emph{iii)} implies that a narrower beam is used for the higher frequency.
Allowing a higher EIRP for the higher band allows an increased coverage area (which becomes similar to that of the lower band), although the drawback is an increased interference to the other cells.

\section{Distributed Cell and Carrier Association}
\label{association}

The network performance under any spectrum access model (exclusive, pooled or hybrid) will depend on 
how a UE is assigned a serving BS and a carrier.  This problem is called cell and carrier association
and can be formalized as follows.
Index the set of all BSs by $i$ and all UEs by $j$.  Let $\mathcal{I}_m$ and
$\mathcal{J}_m$ be the subsets of base stations and UEs for operator $m$\footnote{Note that $\mathcal{J}_m$ represents the set of users for operator $m$, while $J_m$ stands for their number, so $J_m = |\mathcal{J}_m|, \, \forall m \in \mathcal{M}$. The same concept is also used for operators and BSs, thus $I_m = | \mathcal{I}_m|$ and $M = |\mathcal{M}|$.}.
In cell and carrier association, each UE $j$ must be assigned a serving BS cell $i^*(j)$ and a carrier $c^*(j)$.  The selection $(i^*(j),c^*(j))$ will be called the cell-carrier assignment.
Importantly, we assume that the UE can only be assigned a cell from its own operator's network.

Although joint cell and carrier assignment has been discussed extensively in the past, most works in this area have focused on macro/pico user association\cite{assignment1,assignment2,assignment3}, whereas reference~\cite{assignment4} considers user association in multi-carrier settings.
All these works perform a ``one-shot'' optimization where all UEs are reallocated together.

Unlike in these related papers, here we consider the following simple distributed method: each new UE decides an initial cell-carrier assignment when it enters connected mode in the network.
We consider an uncoordinated approach where each UE that joins the network receives status information from the BSs that it can reach and then makes the association decision.
We observe that the proposed distributed schemes do not require any major changes to the signaling procedures of today's systems.
Moreover, they do not require exchange of information among the BSs.
The amount of information exchanged is limited to the load of the carrier's BS and the channel of the link for the two carriers.
The downlink SINR values $\gamma_{ij}^{(c)}$ can be determined from measurement reports exchanged (assuming they account for the beamforming gain) and the load information is received from the pool of candidate base stations.
Then, the cell-carrier assignment is chosen to ensure the desired rate and provide load balancing in the network.

To perform the cell-carrier assignment, we consider two possible greedy heuristics:
\begin{itemize}
\item \textit{Load-aware joint carrier and cell association.} 
In this case, the UE \textit{jointly} selects the serving BS and the carrier so as to maximize its rate without considering the effect of this choice on the other UEs.
Specifically, the UE selects the cell-carrier assignment via the maximization.
\begin{equation}
\left( c^{*}, i^{*} \right)= \argmax_{i \in \mathcal{I}_m, c \in \mathcal {C}} \left( \frac{W^{(c)}}{1 + N^{(c)}_{i}} \log_2\left( 1 + \gamma_{ij}^{(c)}  \right) \right),
\label{max_th_equ}
\end{equation}
where $N_{i}^{(c)}$ is the number of UEs already associated to the $c$-th carrier of the $i$-th base station; $W^{(c)}$ is the bandwidth of the $c$-th carrier, and $\gamma_{ij}^{(c)}$ is the SINR between UE $j$ and BS $i$ if allocated to carrier $c$, given in Eq.~\eqref{equation_sinr} which includes the spatial channel characteristics and beamforming directions.
For pooled carriers, it will also include the interference from other operators in the same band.
We let $\eta_{j}$ denote the resulting maximum rate for the UE:
\begin{equation}
\eta_{j} = \frac{W^{(c^*)}}{1 + N_{i^*}^{(c^*)}} \log_2\left( 1 + \gamma_{i^*j}^{(c^*)} \right).
\end{equation}
When computing the rate $\eta_{j}$, we split the bandwidth among all the users associated to the BS even if during the simulation we allocate the entire bandwidth to a single UE at a time.
Furthermore, with this procedure, the ratio between the total bandwidth $W^{(c)}$ and the number of users $(1 + N^{(c)}_{mi})$ associated to the specific carrier provides the average amount of resources allocated to the $j$-th user over time.

\item \textit{Load-aware carrier-only association.} BS and carrier selection are decoupled, i.e., the UE allocation to the serving BS is kept fixed for the entire simulation, while the UE is allowed to select only the carrier as a function of the load and of the SINR.
That is, the carrier is updated via
\begin{equation}
\left( c^{*} \right)= \argmax_{c \in \mathcal {C}} \left( \frac{W^{(c)}}{1 + N^{(c)}_{i^*}} \log_2\left( 1 + \gamma_{i^*j}^{(c)}  \right) \right),
\label{max_th_equ_new}
\end{equation}
where, instead of trying all the possible BSs $i \in  \mathcal{I}_m$, the algorithm is constrained to keep the current BS $i$, i.e., $i \equiv i^*$, and only optimizes the choice of the carrier $c \in \mathcal{C}$.
\end{itemize}

Both these greedy procedures are computable with the knowledge of load and channel state conditions that are obtained from $N_i^{(c)}$ and $\gamma_{i^*j}^{(c)}$, respectively.

\subsection{Observations and Extensions}
The distributed approach described above results in a lightweight implementation that enables responsiveness to rapid fluctuations of the channel state and traffic conditions.
Nonetheless, relying on distributed rate optimization may lead to sub-optimal solutions.
Conversely, a centralized framework can generate optimal solutions, but the excessive control signaling may affect the responsiveness to channel and traffic dynamics.
The performance gap between distributed approaches and a centralized implementation will be the objective of our future work.

We also note that the approaches in Equations \eqref{max_th_equ} and \eqref{max_th_equ_new} assume a round-robin scheduling.
However, the results can be easily extended to a proportionally fair scheduler, by capturing the effect of different UE rates.

\subsection{Cell and carrier selection simulation}
In the simulation, we use a methodology to evaluate the steady-state behavior of the network. More precisely, we implement an iterative procedure (described in the following) which converges to the long-term load distribution between the two carriers achieved in the hybrid scheme.

In the first phase of the simulation, conditioned on all channel gains, each UE $j \in \mathcal{J}_m$ in the area is associated to the BS $i \in \mathcal{I}_m$ with the highest signal strength.
Note that the choice of BS $i$ is not random but instead based on minimum path loss and so, given the shadowing conditions, deterministic.
After the selection of the best BS, we randomly associate the UE to the BS band at $c_{\ell}$ or $c_{h}$, according to some fixed probabilities $\mathcal{P}_{c_{\ell}}$ and $\mathcal{P}_{c_{h}} = 1 - \mathcal{P}_{c_{\ell}}$. In this study, these initial assignment probabilities are taken equal to $\tfrac{1}{2}$.

\begin{figure}[t!]
\centering
\includegraphics[width=0.88\columnwidth]{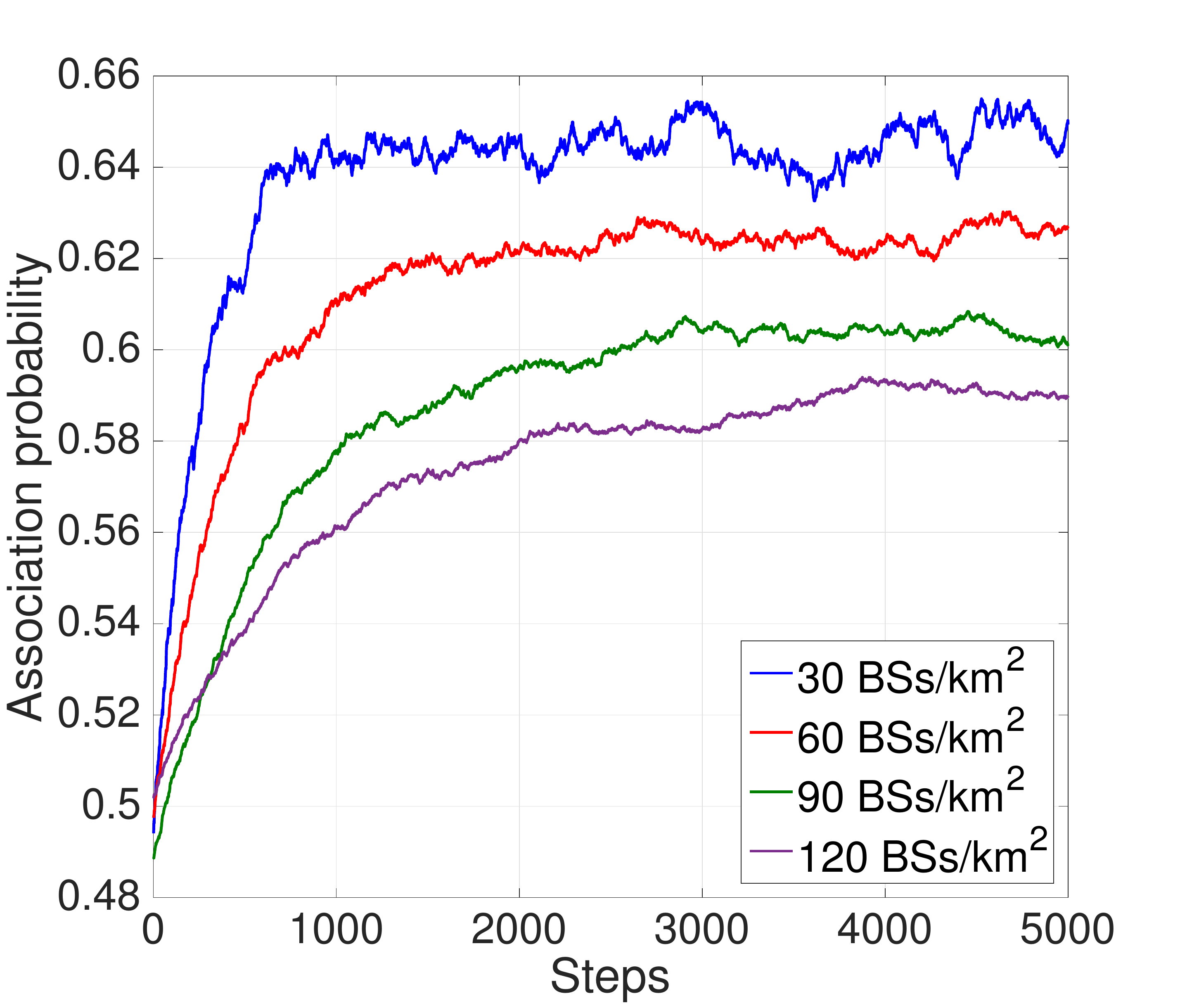}
\caption{Example of the convergence for the probability that a user is associated to carrier $c_{\ell}=$ 28~GHz. Here reported is a run for hybrid case \emph{i)}. The algorithm is initialized with a 0.5 value for each carrier and converges to a stable value which depends on the BS density.}
\label{convergence}
\end{figure}

In the second phase of the simulation, we iteratively update the cell-carrier assignment by  
randomly picking one UE at a time (referred to in the following as UE $j$).  For the selected UE $j$, we update its cell
and carrier using \eqref{max_th_equ} or only its carrier using \eqref{max_th_equ_new}.
We repeat this iterative procedure by re-allocating a random UE at each step, until convergence to a point is reached.
We use the numerical results of Figure~\ref{convergence} to quantify the convergence point and, using this method, we can identify the percentage of users that are connected to $c_{\ell}$ or $c_{h}$.
We note that (cf. Figure \ref{convergence}) the probability that a user is associated to either the $c_{\ell}$ or the $c_{h}$ carrier converges to a stable value, not necessarily equal to the one assumed in the initial phase.
We summarize this iterative procedure in Algorithm~\ref{association_procedure}.

It is important to highlight that the convergence values depend on the different propagation characteristics, bandwidth and amount of interference in the two bands.
Moreover, Figure~\ref{convergence} shows how the number of iterations required  increases as the density of UEs in the area increases.

\begin{figure}[t!]
\centering
\includegraphics[width=0.88\columnwidth]{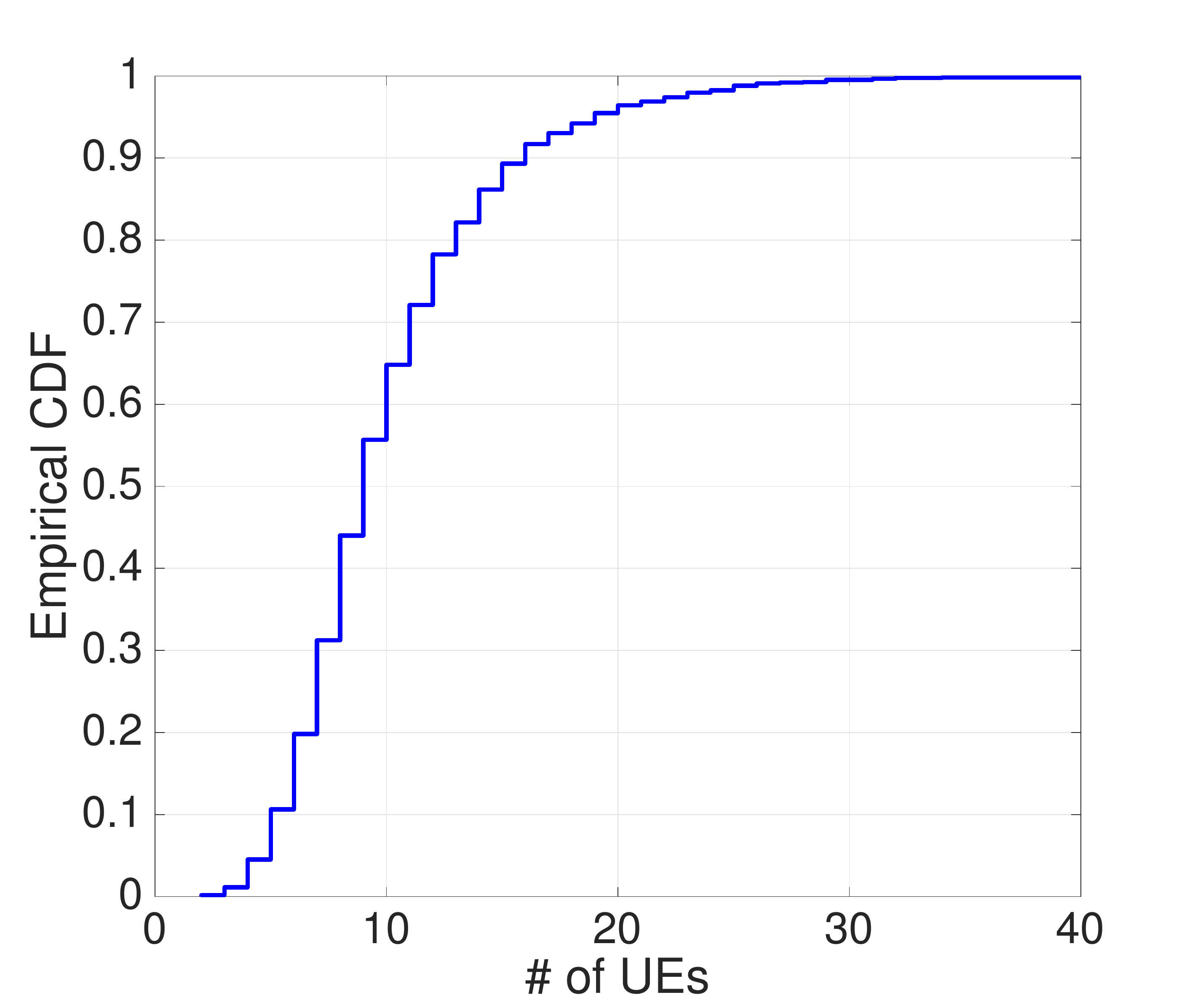}
\caption{CDF of the number of users associated to a single BS.}
\label{ue_distribution}
\end{figure}

Due to the different nature of $i$, it is not possible to provide a convergence plot similar to that for $c^*$.
Therefore, regarding the distribution of UEs per BS, we report in Figure~\ref{ue_distribution} the cumulative density function (CDF) of the number of UEs associated per single BS.
As seen in the figure, the expected number of UEs per BS is close to ten, i.e., equal to the ratio of the two densities. In real conditions, the actual number may be slightly different due to the fact that our model includes random effects such as channel variability, user location and interference.

\begin{algorithm}[!]
\caption{Pseudo-code of the simulation methodology}
\label{association_procedure}
\begin{algorithmic}[1]
\State $\forall m \in \mathcal{M}$ deploy in the area $A$ $J_m$ UEs and $I_m$ BSs following a PPP;
\State $N$: matrix initialized to zeros used to count \# of UEs $\forall i \in \mathcal{I}_m$ and $\forall c \in \mathcal{C}$;
\State $\overline{M}$: vector that stores for each UE the $index$ of the associated BS;
\For {$\forall m \in \mathcal{M}$ and $\forall$ user $j \in \mathcal{J}_m$}
\State Associate user $j$ to the BS $i^*$ with minimum $PL_{mij}$;
\State $p \gets$ randomly pick a value $\in [0,1]$;
\If {$p < \mathcal{P}_{c_{\ell}}$}
\State $c^{*} \gets$ 28~GHz band;
\Else
\State $c^{*} \gets$ 73~GHz band;
\EndIf
\State $N(i^{*}, c^{*}) \gets N(i^{*}, c^{*})  +1$;
\State $\overline{M}(j) \gets (i^{*}, c^{*})$;
\EndFor
\State $P_{\text{TX}_{ic}}$: power set $\forall i \in \mathcal{I}_m$ and $\forall c \in \mathcal{C}$;
\State $G^{(c)}$: computed following Equation~\eqref{max_gains} $\forall c \in \mathcal{C}$;
\State $W^{(c)}$: bandwidth set $\forall c \in \mathcal{C}$;
\State $n$: number of times iterative procedure is repeated;
\For {$n$ times} 
\State $j \gets$ pick random UE in the system;
\State $(i,c) \gets \overline{M}(j)$
\State $N(i, c) \gets N(i, c)  - 1$;
\State $\overline{\gamma}_{mij}^{(c)} \gets$ compute matrix of SINRs $\forall i \in \mathcal{I}_m$, $\forall c \in \mathcal{C}$ as in~\eqref{equation_sinr};
\State $\eta_{mj} \gets$ compute matrix of rates using $\overline{\gamma}_{mij}^{(c)}$, $W^{(c)}$, and $N(i,c)$;
\State $\left( c^{*}, i^{*} \right) \gets \argmax_{i \in \mathcal{I}_m, c \in \mathcal{C}}{(\eta_{mj})}$;
\State $N(i^{*}, c^{*}) \gets N(i^{*}, c^{*})  +1$;
\State $\overline{M}(j) \gets i^{*}$;
\EndFor
\end{algorithmic}
\end{algorithm}

\section{Simulation results}
\label{simulation_results}
%
%

We start by simulating our proposed hybrid spectrum access scheme for the case of joint carrier and cell association.
We consider 4 operators sharing 1~GHz of spectrum at $c_h =$ 73~GHz, while having exclusive access to 250~MHz each at $c_{\ell} =$ 28~GHz (see Fig.~\ref{scenario_draw}).
We also consider two baselines for comparison.
The first is the ``fully licensed'' case where the four operators have each exclusive access to 250~MHz at $c_{\ell}$ and 250~MHz at $c_{h}$.
The second one is the ``fully pooled'' case, where the four operators share 1~GHz at both $c_{\ell}$ and $c_{h}$.
For simplicity, we consider the same UE and BS densities for each operator, although extension to non-homogeneous scenarios would be possible.
All the remaining simulation parameters are reported in Table~\ref{table_of_parameters}.

The results have been averaged over a sufficient number of repetitions in order to obtain the desired accuracy, thus precisely evaluating the proposed hybrid scheme with respect to the baselines. 

\begin{figure*}[h!]
\centering
\begin{subfigure}[b]{0.45\textwidth}
\centering
\includegraphics[width=0.88\textwidth]{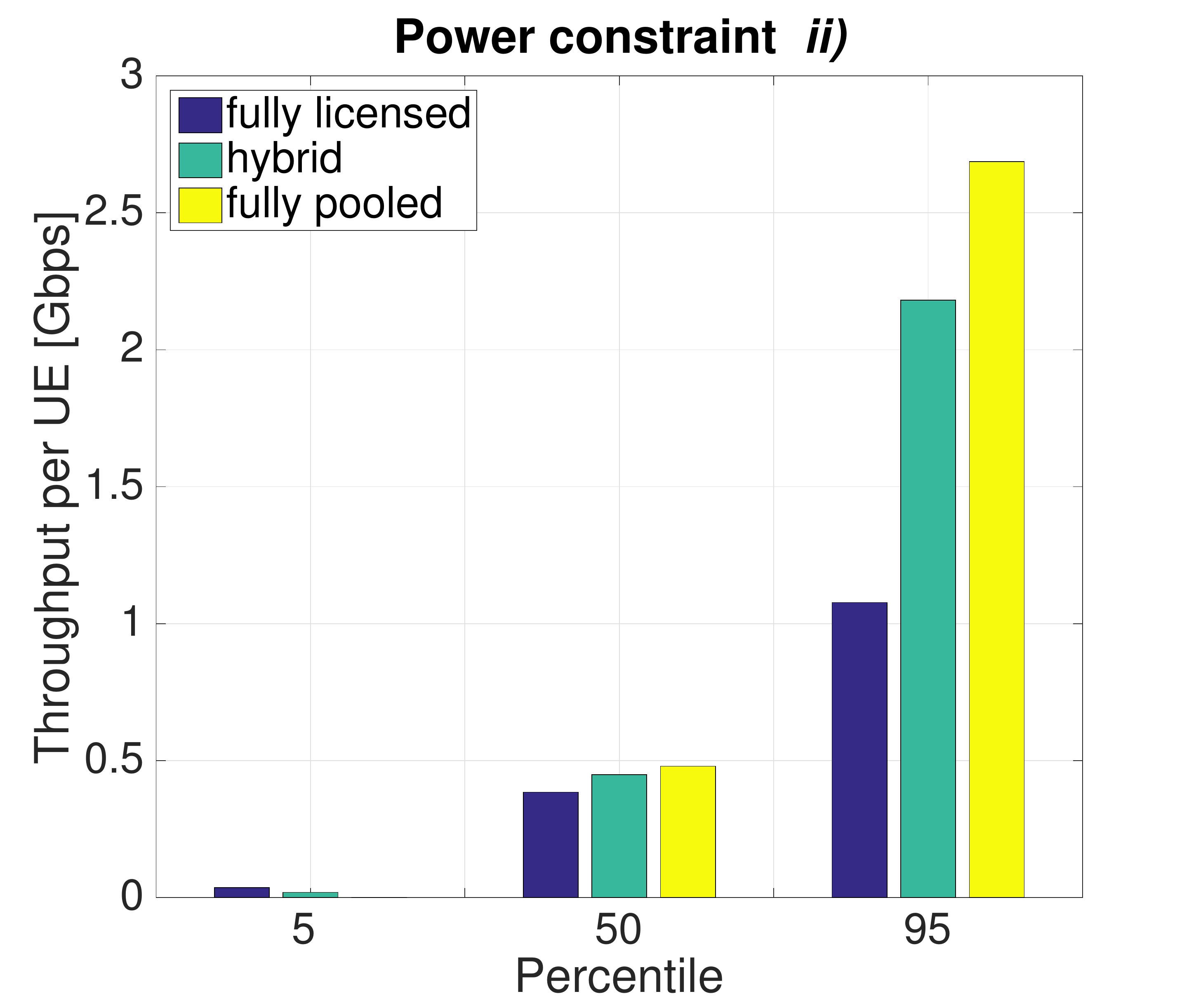}
\caption{Case with BS density equal to 30~BSs/km$^2$.}
\label{fig:throughputs_bar_30}
\end{subfigure}
\hfill
\begin{subfigure}[b]{0.45\textwidth}
\centering
\includegraphics[width=0.88\textwidth]{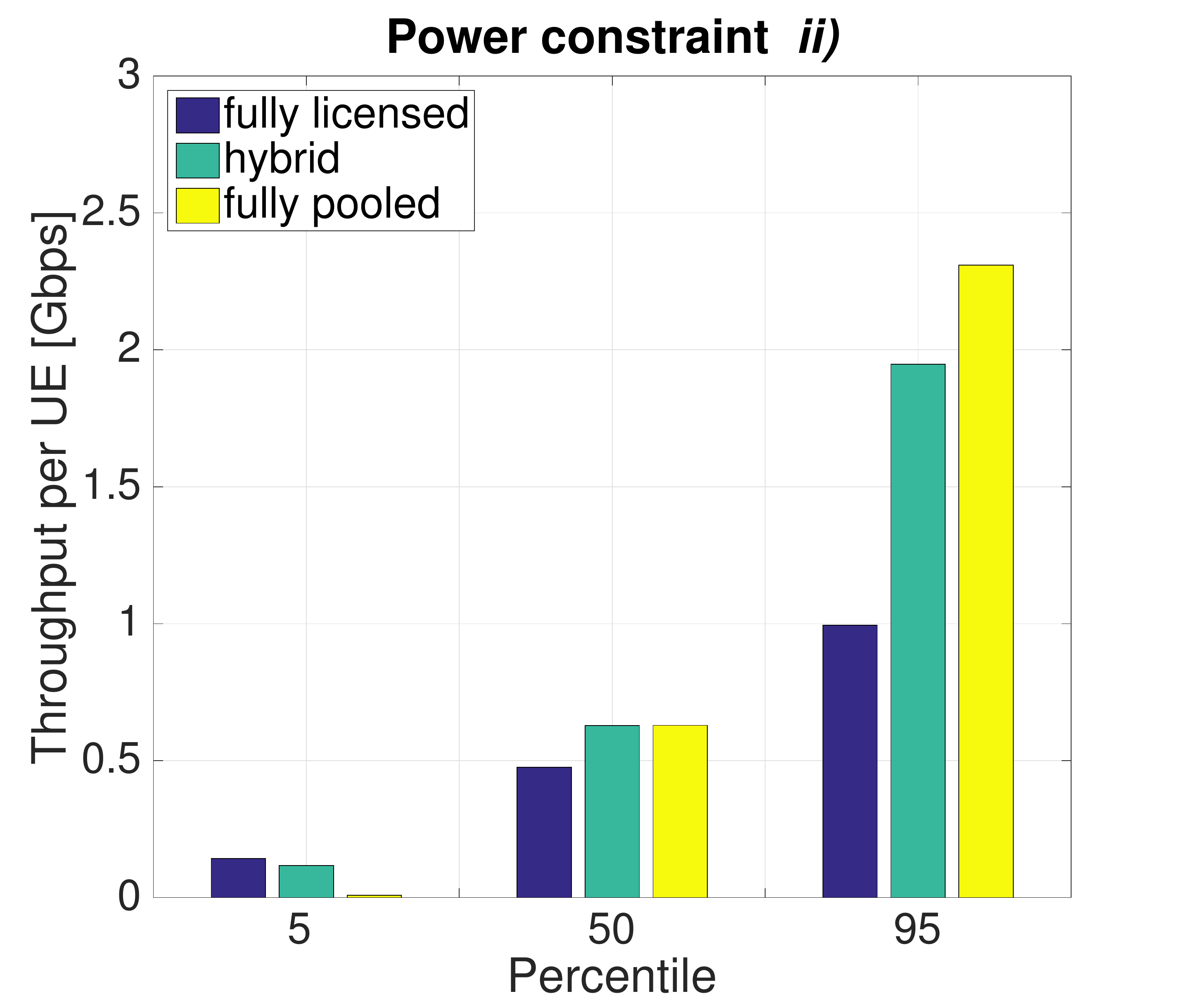}
\caption{Case with BS density equal to 60~BSs/km$^2$.}
\label{fig:throughputs_bar_60}
\end{subfigure}
\caption{Comparison of the throughput measured for the hybrid case and the two baselines. Empirical CDF values of the throughput for the 5th, 50th and 95th percentiles in the power constraint case \emph{ii)}.}
\label{fig:throughputs_bar}
\vspace{-0.3cm}
\end{figure*}

In the first set of results (Figure~\ref{fig:throughputs_bar}), we provide a comparison between the three schemes in terms of 5th, 50th and 95th percentile user rate, for BS densities of 30 and 60~BS/km$^2$. We assume that there are on average ten UEs per BS, so that UE densities of 300 and 600 UEs/km$^2$ have been considered for BS densities of 30 and 60~BSs/km$^2$, respectively.
For these results, we consider the power constraint with an equivalent isotropic radiated power fixed to 48~dBm for both carriers, and we assume twice the number of antenna elements per dimension at 73~GHz with respect to 28~GHz (this model corresponds to power scenario \emph{ii)} above).
\begin{table}[t!]
\normalsize
\renewcommand{\arraystretch}{1.15}
\centering
\begin{tabular}{c|c|c}
\toprule
\textbf{Notation} & \textbf{Value} & \textbf{Description} \\
\hline
\hline
$M$ & 4 & Number of operators\\
\hline
$A$ & 0.3 km$^2$ & Area of simulation\\
\hline
$\lambda_{UE}$& \{300, 600\}  & UE density per km$^2$ \\
\hline
$\lambda_{BS}$& \{30, 60\} & BS density per km$^2$ \\
\hline
$f$ & \{28, 73\}~GHz & Carrier frequency \\
\hline
$P_{\text{TX}}$ & \{24, 30\}~dBm & Transmit power \\
\hline
$W$ & 1~GHz & Total bandwidth\\
\hline
$NF$ & 7~dB & Noise figure\\ 
\bottomrule
\end{tabular}
\caption{Simulation parameters.}
\label{table_of_parameters}
\end{table}

First of all, we can observe how the 95th percentile throughput slightly decreases with the increase of the BS density, since denser topologies result in more interference, thus affecting the performance of the best users. 
This effect is decreased, and in fact almost vanishes, in the fully licensed case due to the reduced number of interfering sources.  
For the 5th and 50th percentiles, increasing the density results in a throughput increase, because the average coverage area is reduced and the BSs are closer to the UEs.
Unlike in the 95th percentile case, here the gain from the closeness of the BS outweighs the loss due to the increased interference.
Especially in the fully pooled case, considering also multiple operators, the amount of interference is very large, and as a result an increased node density does not lead to increased performance.
In fact, the gap between hybrid and fully pooled decreases as the system becomes more interference limited, i.e., denser.

Our proposed hybrid approach also treats the worst users more fairly, as shown by the 5th percentile throughput results in which the hybrid scheme does essentially as well as fully licensed and much better than fully pooled. Conversely, the best users (95th percentile) in the hybrid scheme achieve a throughput that is not much worse than in fully pooled while being significantly better than in fully licensed.

\begin{figure*}[h!]
\centering
\begin{subfigure}[b]{0.45\textwidth}
\centering
\includegraphics[width=0.88\textwidth]{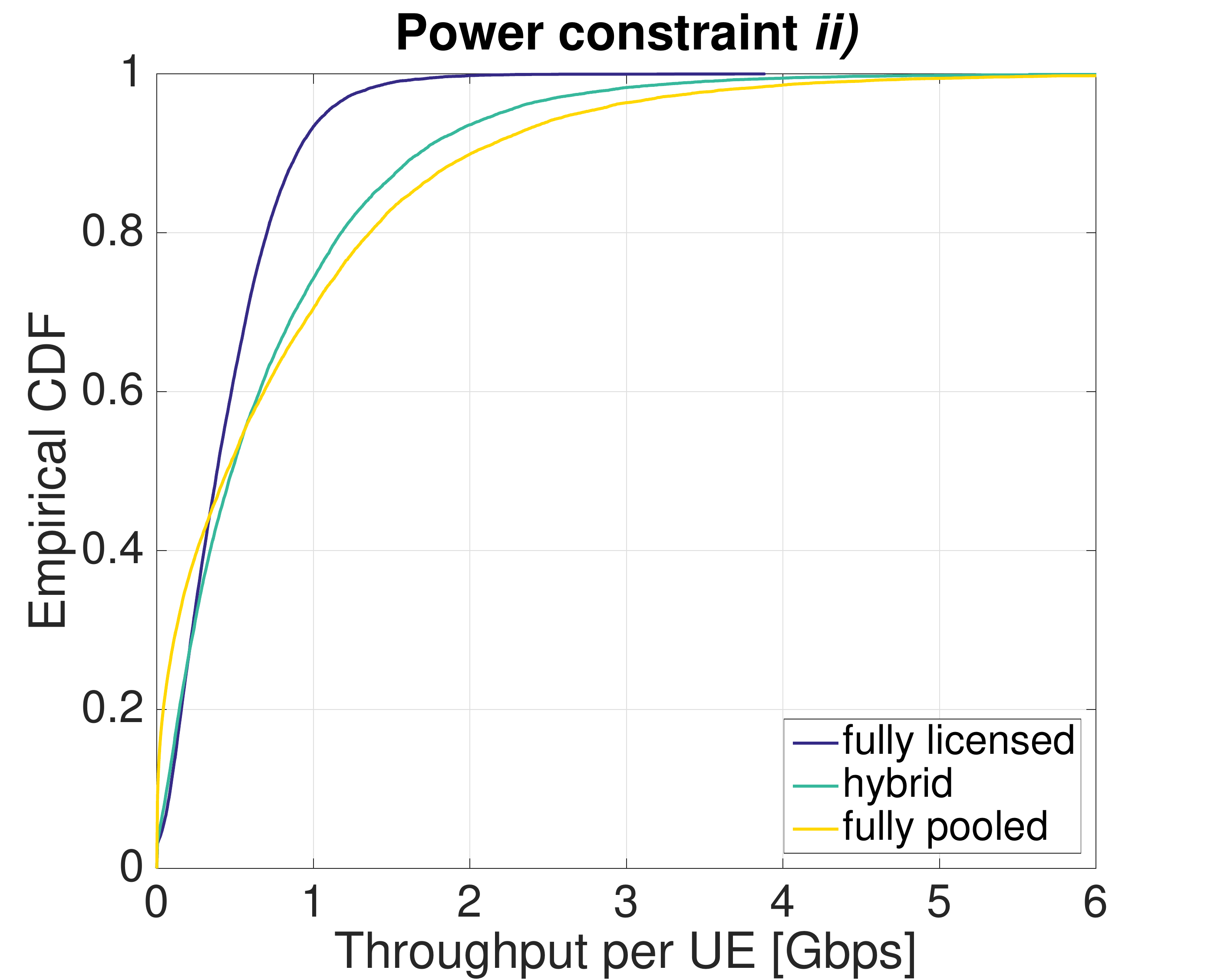}
\caption{Case with BS density equal to 30~BSs/km$^2$.}
\label{fig:throughputs_cdf_30}
\end{subfigure}
\hfill
\begin{subfigure}[b]{0.45\textwidth}
\centering
\includegraphics[width=0.88\textwidth]{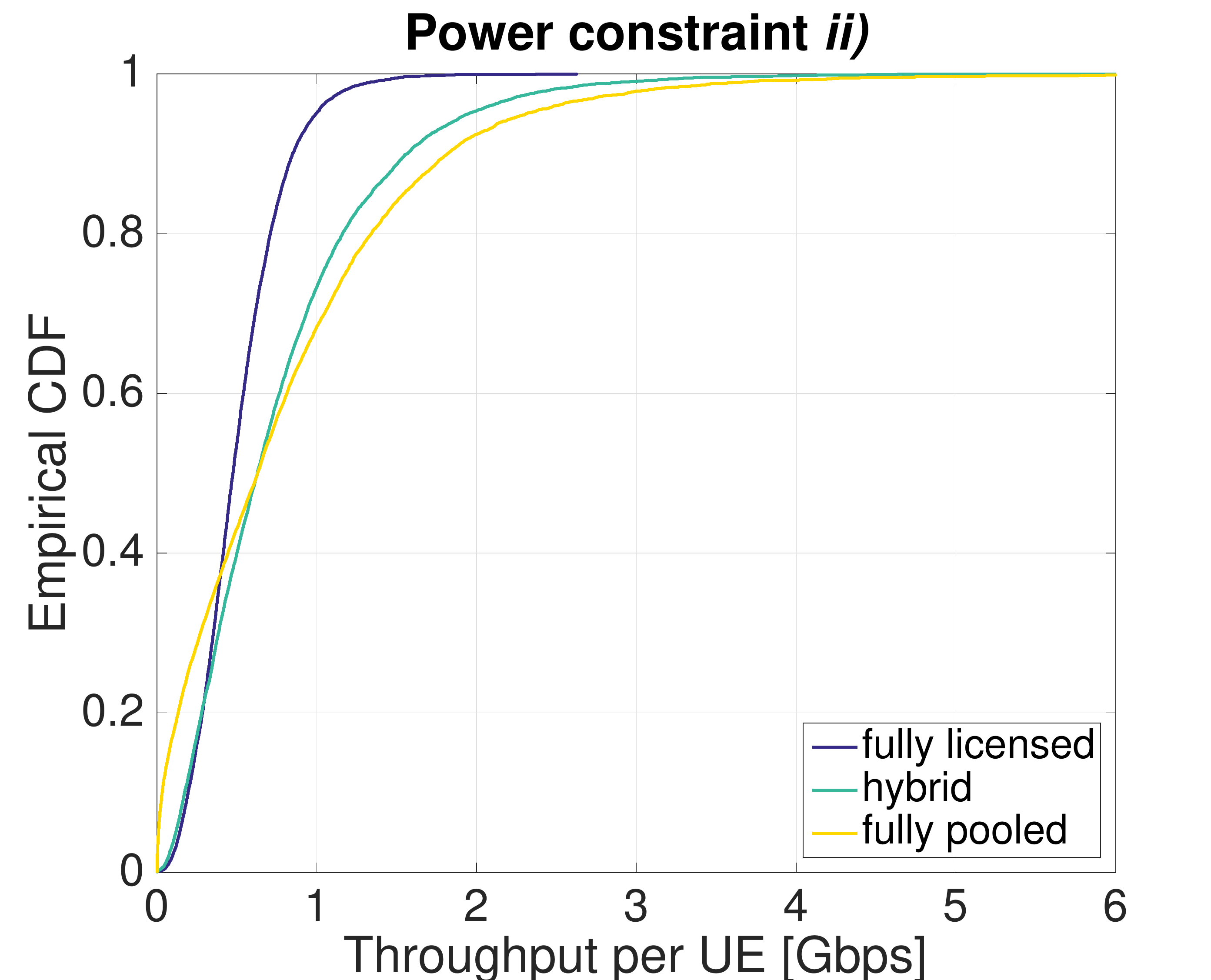}
\caption{Case with BS density equal to 60~BSs/km$^2$.}
\label{fig:throughputs_cdf_60}
\end{subfigure}
\caption{Comparison of the throughput measured for the hybrid case and the two baselines. Empirical CDF curves of the throughput for the power constraint case \emph{ii)}.}
\label{fig:throughputs_cdf}
\vspace{-0.3cm}
\end{figure*}

Comparing the throughput of the worst (5th percentile) users with that of the best (95th percentile) users, we see that they are very different.
This finds an explanation in the fact that, due to directionality, the amount of interference may greatly differ depending on the number and the alignment of the interferers~\cite{rebato17}. 
More detailed performance results are reported in the Empirical CDF plots in Figure~\ref{fig:throughputs_cdf}.
The green curve of the hybrid access is close to the total licensed (blue curve) for the worst users (bottom left), while the curve approaches the total pooled (yellow curve) for the best users (top right), as expected.
This behavior shows that the proposed hybrid scheme is able to closely approach the best performance in various conditions.

In Table~\ref{table_results_30}, which reflects the results shown in Figure~\ref{fig:throughputs_cdf_30}, we provide a comparison for all the different antenna setups and power constraints.
We report the 5th, 50th and 95th percentile values of the user throughput measured in Gbps along with a ratio value, which is computed as the throughput of the hybrid or total pooled cases divided by that of the total licensed case (taken here as the baseline).  It is interesting to note how the throughput varies, at each power constraint.

\begin{table*}[h!]
\normalsize
\renewcommand{\arraystretch}{1.15}
\centering
\begin{tabular}{l|c|c|c|c|cc}
\toprule
                    & \multicolumn{2}{c|}{Power constraint \emph{i)}} & \multicolumn{2}{c|}{Power constraint \emph{ii)}} & \multicolumn{2}{c}{Power constraint \emph{iii)}} \\ \cline{2-7}
                    & Value             & Ratio            & Value              & Ratio            & \multicolumn{1}{c|}{Value}   & Ratio  \\ \hline \hline
Fully licensed 5\%  & 0.0328            & 1.00             & 0.0362             & 1.00             & \multicolumn{1}{c|}{0.0674}  & 1.00   \\ \hline
Hybrid 5\%          & 0.0147            & 0.45             & 0.0190             & 0.52             & \multicolumn{1}{c|}{0.0265}  & 0.39   \\ \hline
Fully pooled 5\%    & 0.0003            & 0.0091             & 0.0007             & 0.019             & \multicolumn{1}{c|}{0.0007}  & 0.010   \\ \midrule
Fully licensed 50\% & 0.3455            & 1.00             & 0.3848             & 1.00             & \multicolumn{1}{c|}{0.4176}  & 1.00   \\ \hline
Hybrid 50\%         & 0.3736            & 1.08             & 0.4492             & 1.17             & \multicolumn{1}{c|}{0.5081}  & 1.22   \\ \hline
Fully pooled 50\%   & 0.3143            & 0.91             & 0.4795             & 1.25             & \multicolumn{1}{c|}{0.4878}  & 1.17   \\ \midrule
Fully licensed 95\% & 0.9188            & 1.00             & 1.0770             & 1.00             & \multicolumn{1}{c|}{1.0218}  & 1.00   \\ \hline
Hybrid 95\%         & 1.9194            & 2.09             & 2.1810             & 2.03             & \multicolumn{1}{c|}{2.0252}  & 1.98   \\ \hline
Fully pooled 95\%   & 2.1926            & 2.39             & 2.6970             & 2.49             & \multicolumn{1}{c|}{2.5461}  & 2.49   \\
 \bottomrule
\end{tabular}
\caption{Values of the throughput (measured in Gbps) for the hybrid case and the two baselines. Rates for the 5th, 50th and 95th percentiles and all the power constraints, simulations with a BS density of 30~BSs/km$^2$. The ratio values are computed as the throughput of the hybrid or total pooled cases divided by that of the total licensed case, which is taken as a reference.}
\label{table_results_30}
\end{table*}

We recall from Section \ref{simulation_methodology} that in case $i)$ we consider the same number of antenna elements at 28~GHz and 73~GHz, in case $ii)$ we double the number of antenna elements per dimension at 73~GHz with an equal constraint on the EIRP, and, finally, in case $iii)$ we use the configuration of case $ii)$ but with different EIRP constraints for the two bands.
Constraint \emph{ii)} represents a trade-off for the available throughput between the number of antenna elements usable for a fixed area and the power radiated in the environment.

From Table~\ref{table_results_30}, we can note that a large number of antenna elements has a positive impact on the 95th percentile user rate for both the hybrid and the fully pooled case.
This relates to coverage, along with the amount of interference generated within the cell: using more antennas results in narrower beams, and hence reduced interference, which in turn leads to higher throughput.
Further, by looking at the ratio values reported in Table~\ref{table_results_30}, we can better understand the performance gains obtained with a hybrid scheme vs. a total pooled spectrum access.
In the more realistic scenario, i.e., power constraint \emph{ii)}, the 5\% user throughput of the hybrid case is smaller than that obtained via a \emph{fully licensed} scheme (about 50\% smaller), and on the other hand much higher than the rate achieved with a \emph{total pooled} approach (about 32 times higher).
A similar behavior is observed for all the power configurations at the 5th percentile, while the opposite occurs if we consider the 95\% user throughput.
This last behavior is less pronounced and for this reason we can conclude that our hybrid procedure exhibits desirable performance, and represents a tradeoff between the two baselines. This is due to the fairness that originates from our opportunistic carrier selection.

\begin{figure}[t!]
\centering
\includegraphics[width=0.88\columnwidth]{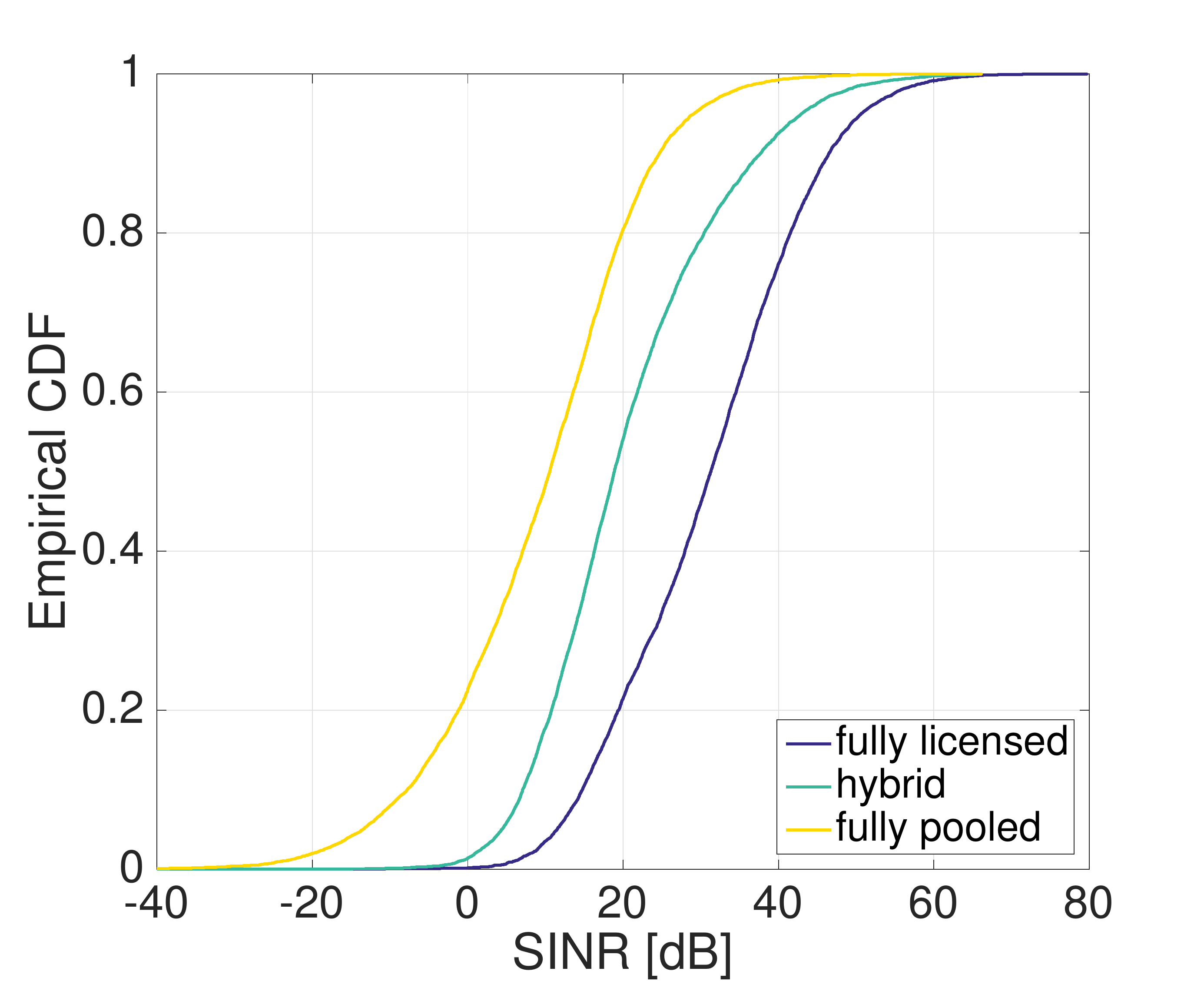}
\caption{Comparison of the SINR measured for the hybrid case and the two baselines. Empirical CDF for the power constraint case \emph{ii)} and with BS density equal to 60~BSs/km$^2$.}
\label{fig:sinr_cdf_60}
\end{figure}

We also show in Figure~\ref{fig:sinr_cdf_60} the Empirical CDF of the SINR for the hybrid spectrum access scheme and the two baselines fully licensed and fully pooled. 
When bandwidth sharing is increased, the average SINR decreases and the CDF curve moves to the left.
However, it is important to highlight that our approach does not have the goal of optimizing the SINR but rather tries to maximize the user rate, considering also the load in terms of UEs connected to the BS, which is not considered in Fig.~\ref{fig:sinr_cdf_60}.

\begin{figure*}[t!]
\centering
\includegraphics[width=0.88\textwidth]{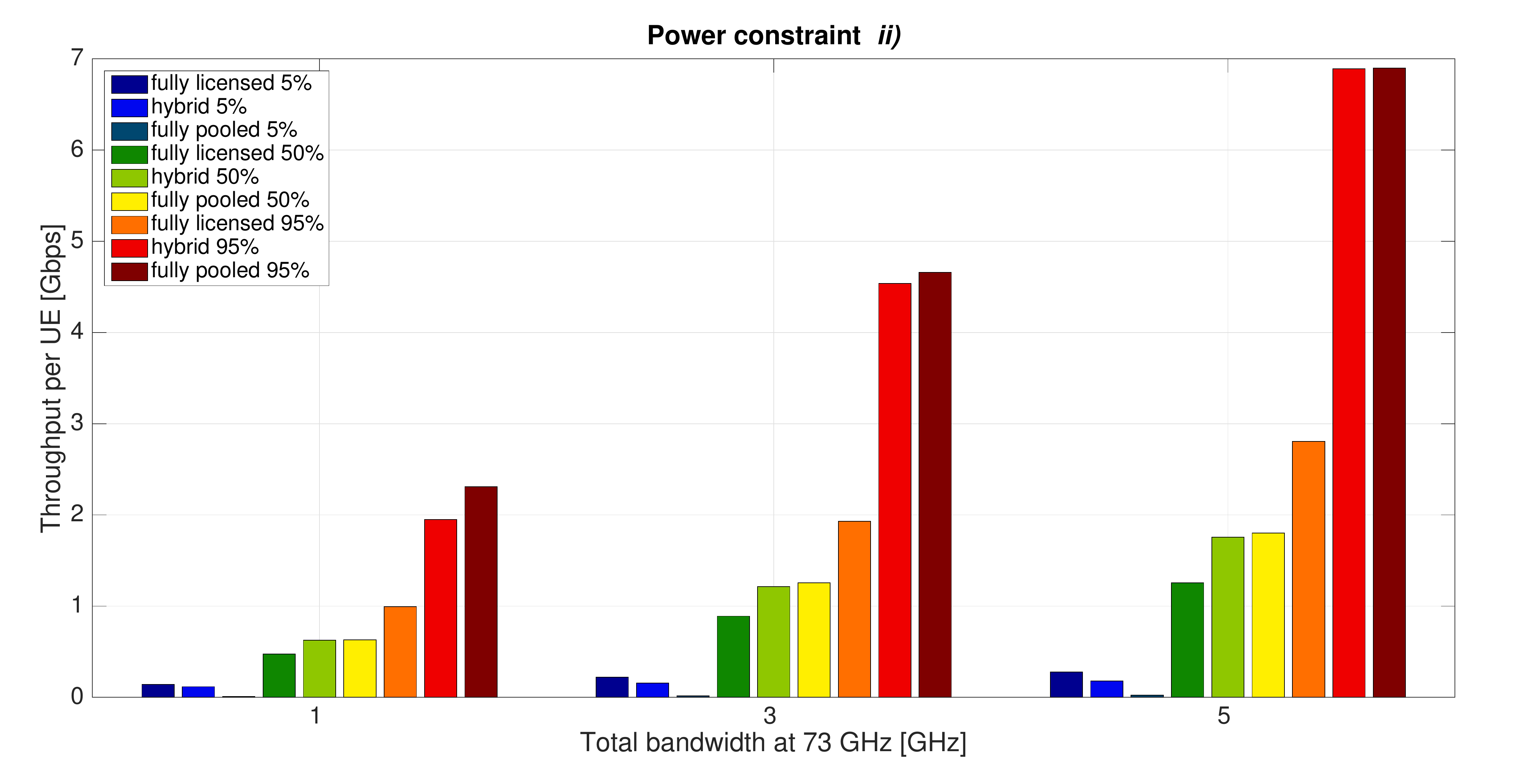}
\caption{Comparison of the average throughput measured for the hybrid case and the two baselines varying the total bandwidth assigned to the 73~GHz band. Simulations with a BS density of 60~BSs/km$^2$.}
\label{fig:bandwidth_ii_th}
\vspace{-0.4cm}
\end{figure*}

We present a further evaluation in Figure \ref{fig:bandwidth_ii_th}, aiming to  assess the performance impact of different available bandwidths (1, 3 and 5~GHz) for the $c_{h}$ frequency carrier.
Once again, the hybrid scheme exhibits similar trends as in the fully licensed case for the worst users, while approaching the fully pooled case performance gains for the best users.
Even if the thermal noise and the interference increase with larger bandwidth, improvements in the throughput are observed in each scenario when the band increases. 
Moreover, the gap between the hybrid and total pooled schemes is reduced when the total bandwidth used at 73~GHz is increased.

\begin{figure*}[t!]
\centering
\begin{subfigure}[b]{0.45\textwidth}
\includegraphics[width=0.88\textwidth]{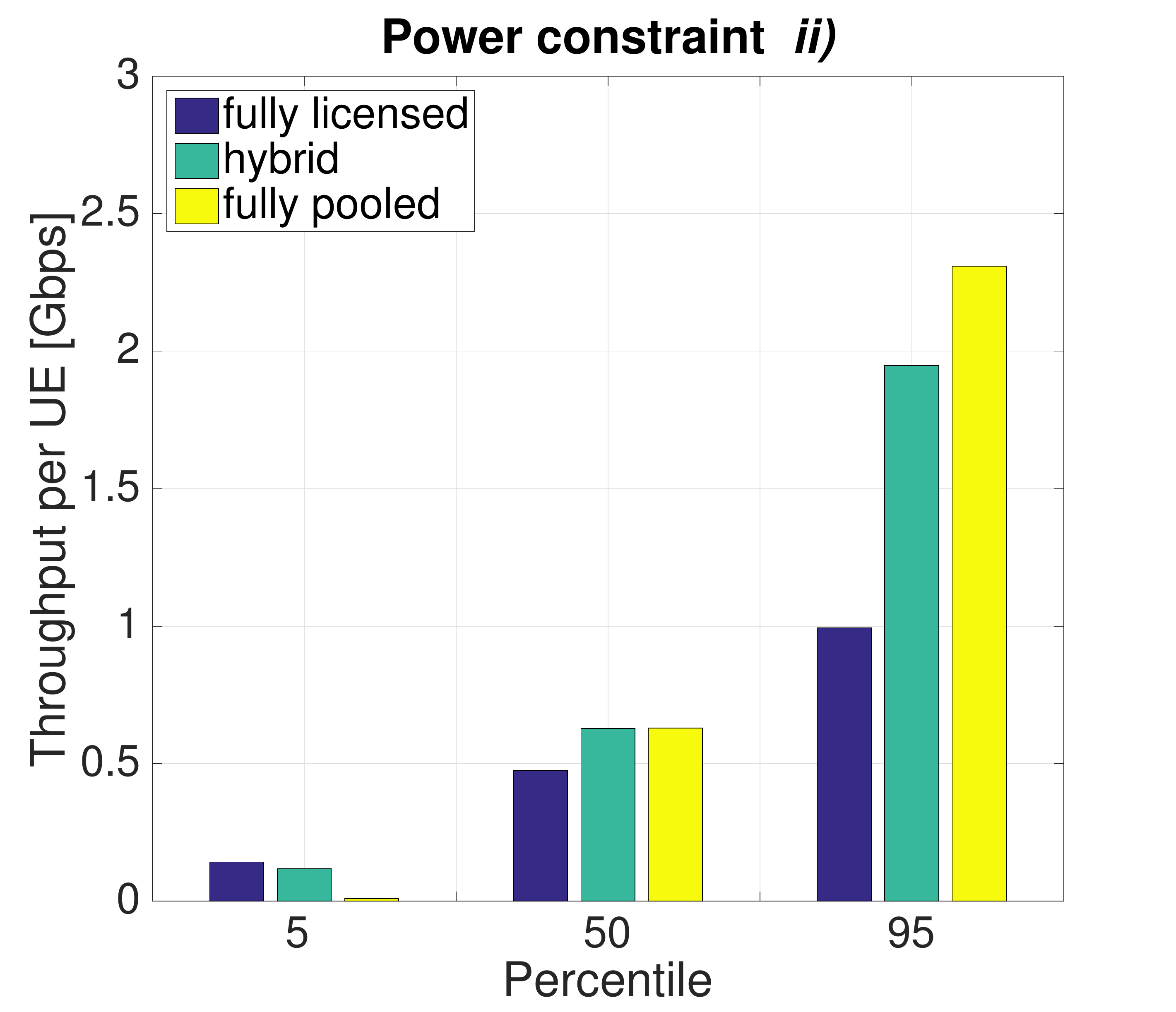}
\caption{Load-aware joint carrier and cell association.}
\label{bar_oc_normale_case}
\end{subfigure}
\hfill
\begin{subfigure}[b]{0.45\textwidth}
\includegraphics[width=0.88\textwidth]{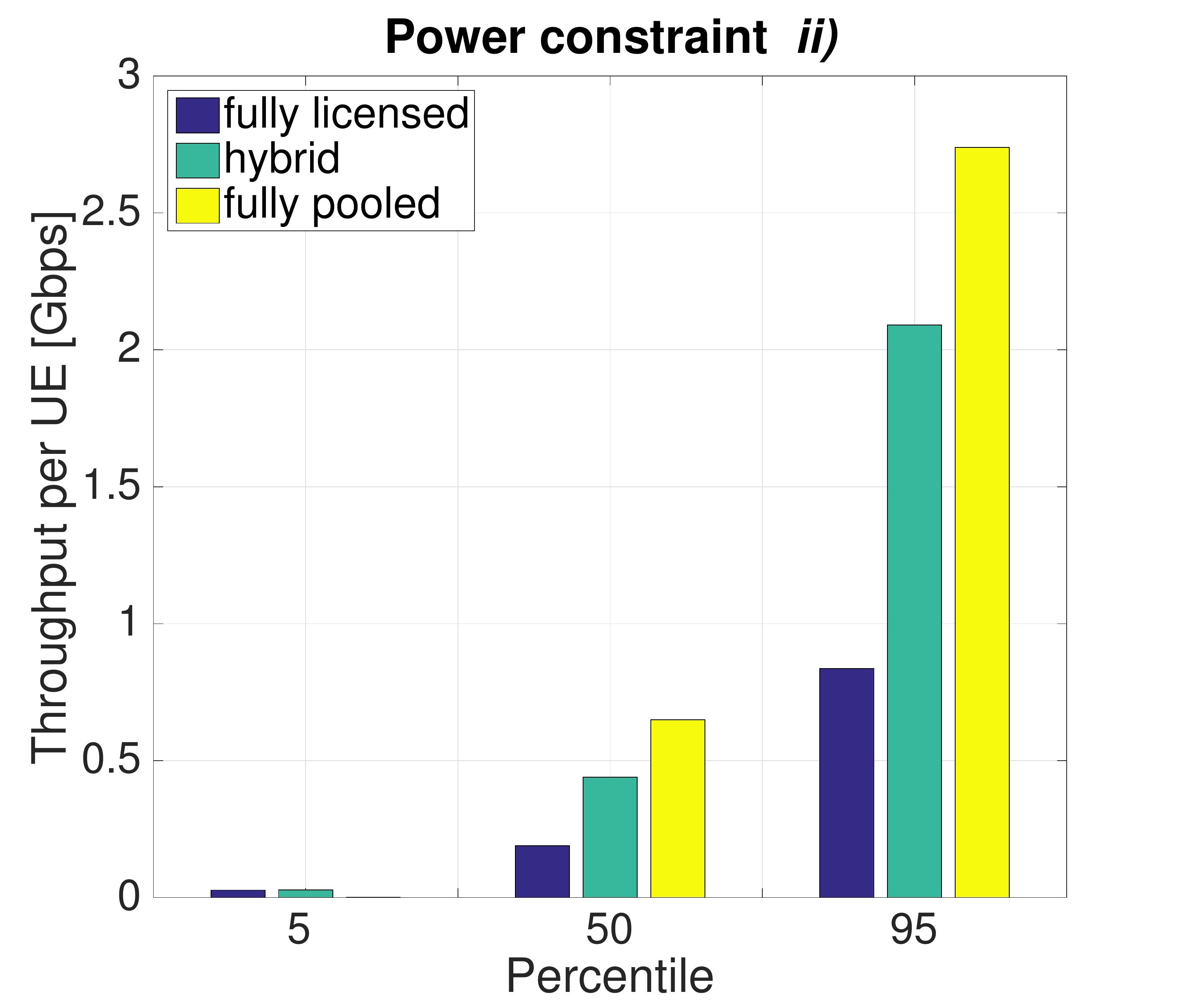}
\caption{Load-aware carrier association.}
\label{fig:bar_oc_60}
\end{subfigure}
\caption{Comparison of the throughput measured for the hybrid case and the two baselines and for the two different association algorithms. Values of simulations of the power constraint \emph{ii)} with a BS density of 60~BSs/km$^2$.}
\label{fig:bar_oc}
\vspace{-0.4cm}
\end{figure*}

Finally, we report in Figure \ref{fig:bar_oc} a comparison between the different association algorithms, where each UE either jointly selects the serving BS and the frequency band (left plot), or can only choose the optimal carrier while keeping the BS association fixed (right plot).
From these results, we can observe that the throughput provided by \emph{carrier-only} association is higher than by \emph{joint} association for the best users (95th percentile), while the opposite occurs for the median and worst (5th percentile) users.
This can be explained by observing that in the carrier-only association it is not possible to redistribute users to balance the load among BSs, and therefore the best users are likely to be those that happen to be associated to lightly loaded BSs, enjoying a higher rate compared to what they would achieve under the joint cell and carrier association. For the same reason, the worst users (likely associated to highly loaded BSs) experience very poor performance. On the contrary, joint cell and carrier association tends to distribute users among BSs and carriers more equitably, and results in a fairer system.


\section{Conclusion and future works}
\label{futureworks}

In this paper, we have introduced a new hybrid spectrum access scheme for mmWave networks, where data packets are scheduled through two mmWave carriers with different characteristics. In particular, we have proposed a spectrum sharing scheme which combines a lower mmWave band with exclusive access and a higher mmWave band where spectrum is pooled between multiple operators.
Our investigation shows that this approach provides advantages for the average user with respect to traditional fully licensed or fully pooled spectrum access schemes, in terms of increased throughput and spectral efficiency.
The approach offers also better balancing of the available resources with respect to the fully pooled case, resulting in higher fairness.
This work opens a promising line of research towards a more flexible and efficient use of the radio spectrum, and will provide useful input and insights to standardization and spectrum policy.

However, further work is needed to reach a more complete understanding of different aspects.  First, the study in this paper considers the case where the higher mmWave band is pooled between multiple operators all having a licensed anchor in another mmWave band at a lower frequency. It would be interesting to study a more general spectrum arrangement where the higher mmWave band is entirely license-exempt, i.e., can be accessed by heterogeneous users with and without an anchor at a lower frequency. The main challenge in this case is how to design politeness mechanisms in a very directional propagation environment.
On one side traditional listen-before-talk techniques would not provide a reliable solution.
On the other side, more recent directional MAC approaches introduced for 802.11ad (that is based on fully unlicensed spectrum access) would be suboptimal in a hybrid mmWave spectrum context.
Second, our study is based on a distributed uncoordinated algorithm.
We chose this approach because it allows an initial assessment based on a practical mechanism that has a low impact on signaling and architecture design.
On the other hand, we believe that centralized approaches could also provide a realistic solution, in particular in networks where infrastructure sharing is used.
Third, while this study provides an initial, proof-of-concept assessment based on simulation, we believe that a mathematical analysis could lead to a more fundamental understanding of the different factors underlying hybrid spectrum access.
Finally, it would also be interesting to evaluate our cell and carrier selection methods under time-varying traffic, where users come and go according to some statistics. In this case, the dynamics of traffic and interference will interact with the user allocation strategies and will accordingly lead to a time-varying throughput performance, whose characterization is an interesting item of future study.

\section*{Acknowledgment}
The work of F. Boccardi for
this paper was carried out in his personal capacity and the
views expressed here are his own and do not reflect those of
his employer

\bibliographystyle{IEEEtran}
\bibliography{biblio}
\end{document}